\documentclass[aps,prl,twocolumn,preprintnumbers,amsmath,amssymb,10pt]{revtex4-2}

\usepackage[utf8]{inputenc}
\usepackage{graphicx}
\usepackage{bm}
\usepackage{amsmath}

\usepackage{color}
\definecolor{darkblue}{rgb}{0,0,0.6}
\definecolor{red}{rgb}{1,0,0}

\usepackage{ifpdf}
\ifpdf
\usepackage{epstopdf}
\usepackage[pdftex,unicode,pdfstartview={FitH},pdfborder={0 0 0}]{hyperref}
\usepackage{hypcap}
\else
\usepackage[hypertex]{hyperref}
\fi
\hypersetup{
    bookmarksnumbered = true,
    colorlinks = true, linkcolor = darkblue,
    citecolor = darkblue, filecolor = darkblue,
    menucolor = darkblue, urlcolor = darkblue
}

\usepackage{dcolumn}
\newcolumntype{R}{>{$\displaystyle}r<{$}}
\newcolumntype{C}{>{$\displaystyle}c<{$}}
\newcolumntype{L}{>{$\displaystyle}l<{$}}

\begin{document}
    
\title{Correlated magnetism of moir\'e exciton-polaritons on a triangular electron-spin lattice}

\author{Johannes Scherzer$^{1,*}$, Lukas Lackner$^{2,*}$, Bo Han$^{2}$, Borislav Polovnikov$^{1}$ , Lukas Husel$^{1}$, Jonas G\"oser$^{1}$, Zhijie Li$^{1}$, Jens-Christian Drawer$^{2}$, Martin Esmann$^{2}$, Christoph Bennenhei$^{2}$, Falk Eilenberger$^{3,4}$, Kenji Watanabe$^{5}$, Takashi Taniguchi$^{6}$, Anvar S. Baimuratov$^{1}$, Christian Schneider$^{2}$, Alexander H\"ogele$^{1,7}$}

    \affiliation{$^1$Fakult\"at f\"ur Physik, Munich Quantum Center, and Center for NanoScience (CeNS), Ludwig-Maximilians-Universit\"at M\"unchen, Geschwister-Scholl-Platz 1, 80539 M\"unchen, Germany}

    \affiliation{$^2$Institute of Physics, Carl von Ossietzky Universit\"at Oldenburg, Carl-von-Ossietzky-Straße 9-11, 26129 Oldenburg, Germany}
    
    \affiliation{$^3$Institute of Applied Physics, Friedrich-Schiller-Universit\"at Jena, Max-Wien-Platz 1, 07743 Jena, Germany}    
    
    \affiliation{$^4$Fraunhofer-Institute of Applied Optics and Precision Engineering IOF, Albert-Einstein-Straße 7, 07745 Jena, Germany}
    
    \affiliation{$^5$Research Center for Electronic and Optical Materials, National Institute for Materials Science, 1-1 Namiki, Tsukuba 305-0044, Japan}
    
    \affiliation{$^6$Research Center for Materials Nanoarchitectonics, National Institute for Materials Science,  1-1 Namiki, Tsukuba 305-0044, Japan}
       
    \affiliation{$^7$Munich Center for Quantum Science and Technology (MCQST), Schellingstra\ss{}e 4, 80799 M\"unchen, Germany}

    \affiliation{$^*$These authors contributed equally to this work.}

\maketitle

\textbf{Moir\'e van der Waals materials support a plethora of correlated many-body phenomena, including single-lattice  \cite{WuHubbard2018} and bilayer  \cite{Polovnikov2024} Hubbard model physics, Mott insulating states of generalized Wigner crystals \cite{Xu2020,Regan2020,Shimazaki2020}, or quantum anomalous Hall phases \cite{Serlin2020,LiMak2021,Cai2023,Park2023,Xu2023}. Moreover, they give rise to emergent magnetism on triangular electron \cite{Ciorciaro2023} or hole \cite{Tang_2020,Xu2022,Campbell_2022,Tang_2023} spin lattices with nonlinear moir\'e exciton Land\'e factors as peculiar signatures. Here, we provide evidence of correlated magnetism phenomena for exciton-polaritons in a MoSe$_{2}$/WS$_{2}$ van der Waals heterostructure with near-parallel alignment subject to electron doping. In our experiments, interactions between electrons and moir\'e excitons are controlled electrostatically by field-effect doping \cite{Ciorciaro2023,Polovnikov_2023,Polovnikov2024}, and the polaritonic regime of strong light-matter coupling is established in an open cryogenic microcavity \cite{Schneider_2018,Drawer_2023}. Remarkably, at filling fractions around one electron per moir\'e cite, we observe drastic and nonlinear enhancement of the effective polariton Land\'e factor as a hallmark of correlated magnetism, which is cavity-controlled via resonance tuning of light and matter polariton constituents. Our work establishes moir\'e van der Waals heterostructures as an outstanding platform for studies of correlated phenomena in the presence of strong light-matter coupling and many-body phases of lattice-ordered charges and spins.}

Exciton-polaritons are bosonic quasi-particles arising in the regime of strong coupling between photons and excitons. They become particularly distinct in optical micro-cavities coupled to materials with excitons of significant oscillator strength \cite{Weisbuch_1993, Kavokin_2011}. Being composite bosons, they possess a variety of unique properties resulting from light-matter hybridization on a fundamental quantum level, including giant optical nonlinearities that give rise to effective photon-photon interactions \cite{MunozMatutano_2019, Delteil_2019}, the formation of quantum condensates at elevated densities from cryogenic to room temperatures \cite{Kasprzak_2006,Anton-Solanas_2021,Kena-Cohen_2010}, as well as friction-less on-chip propagation \cite{Amo_2011}. Complementary to conventional semiconductors, van der Waals systems of transition metal dichalcogenides (TMDs) and their heterostructures provide novel opportunities for engineering quantum phases in the regime of strong light-matter coupling \cite{Schneider_2018}. 

In TMD monolayers, reduced screening enhances nonlinear polariton interactions \cite{Gu2021}, while controlled electron doping enables the formation of Fermi polaron-polaritons by dressing exciton-polaritons with excitations in the Fermi sea \cite{Sidler_2017,Efimkin_2017}. Remarkably, the paramagnetic behavior of such charged light-matter quasiparticles is dominated by the spin-polarization of the Fermi sea \cite{Back_2017}, rendering polaron-polaritons sensitive to external magnetic fields \cite{Lyons_2022}. In TMD heterobilayer systems, on the other hand, the emergent moir\'e potential leads to the formation of electron or hole lattices at specific fractional fillings of elementary charges in units of the moir\'e density. These lattices exhibit correlated magnetic phases, which can be controlled with an external magnetic field and detected through the optical response of moir\'e excitons \cite{Tang_2020,Xu2022,Campbell_2022,Tang_2023,Morera_2023,Ciorciaro2023,Polovnikov2024}. Dressing such moir\'e exciton polarons with cavity photons would allow the exploration and control of the nonlinear manifestations of correlated magnetism for polaron-polaritons as an extension to nonlinear optical phenomena observed for charge-neutral moir\'e exciton-polaritons \cite{Zhang_2021}.

Here, we employ the triangular moir\'e lattice of a MoSe$_{2}$/WS$_{2}$ heterostructure with near-parallel alignment, and demonstrate how correlated magnetism of electrons maps onto the magnetic response of neutral moir\'e exciton-polaritons and charged polaron-polaritons. In our experiments, we combine field-effect control of electron density in a dual-gate device with a fully tunable optical microcavity in a magneto-optical cryostat. Remarkably, charged moir\'e exciton-polaritons reveal a highly nonlinear response to out-of-plane magnetic fields as a hallmark of correlation-induced magnetism, most pronounced near the Mott insulating phase at unity electron filling factor ($\nu=1$), where kinetic magnetism supports ferromagnetic correlations \cite{Ciorciaro2023,Morera_2023}. In this regime, moir\'e polaron-polaritons uniquely provide access to cavity-controlled effective Land\'e factors beyond the usual paramagnetic response of bare excitons \cite{Back_2017} and monolayer polaron-polaritons \cite{Lyons_2022}.

\begin{figure}[t]  
\includegraphics[width=\columnwidth]{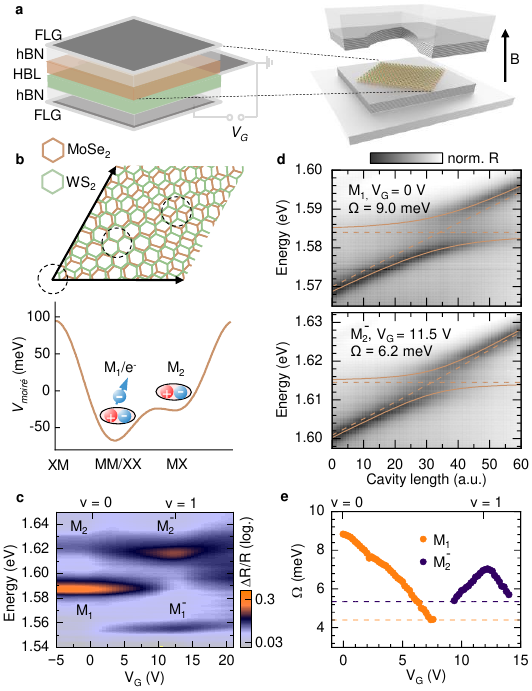}
\caption{\textbf{Moir\'e exciton-polaritons and polaron-polaritons in a gate-tunable MoSe$_{2}$/WS$_{2}$ heterostructure.} \textbf{a}, Schematic of the heterostructure device, placed in a microcavity inside a closed-cycle cryostat with out-of-plane magnetic field B. \textbf{b}, Unit cell of the moir\'e lattice with high-symmetry points (dashed circles) and the corresponding moir\'e exciton potential. \textbf{c}, Electron charging in reflection contrast spectroscopy of moir\'e excitons M$_{1}$ and M$_{2}$ and the respective moir\'e polarons M$_{1}^-$ and M$_{2}^-$ as a function of the gate voltage V$_{G}$. \textbf{d}, Characteristic anticrossing of M$_{1}$ exciton-polariton (upper panel) and M$_{2}^{-}$ polaron polariton (lower panel) branches as a function of cavity-resonance detuning (the model fits are shown by solid lines, and the bare exciton and cavity dispersions by dashed lines). \textbf{e}, Rabi splitting $\Omega$ of M$_{1}$ and M$_{2}^{-}$ polaritons obtained from the coupled oscillator model as a function of V$_{G}$. Horizontal dashed lines delimit the strong coupling condition $\Omega> (\gamma+\kappa)/2$.}
\label{fig1}
\end{figure}

The dual-gate device is shown schematically in Fig.~\ref{fig1}a. The MoSe$_{2}$/WS$_{2}$ heterobilayer (HBL) with a small angle (below $2$°) away from parallel alignment was encapsulated in hexagonal boron nitride (hBN) and sandwiched between few-layer graphite (FLG) electrodes for doping and electric field control. The van der Waals heterostructure was integrated in a tunable open microcavity formed by distributed Bragg reflector (DBR) mirrors and mounted inside a low vibration closed-cycle cryostat with a solenoid providing out-of-plane magnetic fields \cite{Drawer_2023} (see the Methods section for details). The TMD monolayers were synthesized by chemical vapor deposition \cite{Bilgin_2015}, featuring triangular crystal shapes that enable precise control of their relative orientation in an optical microscope during assembly. The twist angle and lattice mismatch between the two TMD layers gives rise to a moir\'e superlattice (Fig.~\ref{fig1}b, upper panel) which modulates the energy of both electrons and excitons \cite{Wu_2017,WuHubbard2018,Wu_2018_jan,Tong_2020} throughout the moir\'e unit cell (Fig.~\ref{fig1}b, lower panel).

First, we establish the origin of moir\'e exciton and polaron resonances in the heterostructure with white light reflection contrast spectroscopy in the absence of the top DBR mirror. The evolution of the exciton oscillator strength as a function of the electron density, controlled by the voltage V$_{G}$ on the bottom gate of the heterostack, is shown in Fig.~\ref{fig1}c. At charge neutrality (V$_{G}=0$~V), the optical response displays two prominent features M$_{1}$ and M$_{2}$ at $1.584$ and $1.620$~eV, respectively, in the spectral vicinity of the fundamental exciton transition in monolayer MoSe$_{2}$. We assign M$_{1}$ to the renormalized MoSe$_{2}$ ground state exciton, and M$_{2}$ to the first umklapp moir\'e exciton band \cite{Alexeev_2019}. The absence of a pronounced linear Stark shift in an out-of-plane electric field (Supplementary Note~1), the prominent oscillator strength, and the narrow linewidth of both exciton transitions suggest a type-I band alignment, yielding lowest energy moir\'e excitons of predominant MoSe$_{2}$ intralayer character \cite{Tang_2021,Tang_2022,Polovnikov2022}. The combined features indicate very weak degree of interlayer hybridization for both M$_{1}$ and M$_{2}$ moir\'e exciton states, with wavefunctions mainly localized in the MoSe$_2$ layer \cite{Polovnikov_2023}.

Within the specific moir\'e potential of near-parallel MoSe$_{2}$/WS$_{2}$ heteorstacks \citep{Ciorciaro2023}, electrons and M$_{1}$ excitons preferably localize at moir\'e lattice sites of aligned metal atoms (MM), while M$_{2}$ has its energetically favored position at MX sites (see Fig.~\ref{fig1}b and Supplementary Note~1). With increasing electron density, two new resonances, M$_{1}^{-}$ and M$_{2}^{-}$, emerge in the regime of electron filling per moir\'e cell $0<\nu<1$, and acquire sizable oscillator strength at the expense of M$_{1}$ and M$_{2}$. This behavior is reminiscent of excitons interacting with electrons in the Fermi sea \cite{Back_2017,Sidler_2017,Efimkin_2017}, forming correlated electron-exciton states. Whereas M$_{1}^-$ attractive polarons are formed as bound states of excitons and electrons on the same lattice site with a binding energy comparable to mononlayer MoSe$_2$ \cite{Ross_2013,Sidler_2017}, M$_{2}$ excitons are localized on a distant lattice site, which results only in a weak redshift of the M$_{2}^{-}$ attractive polaron resonance. Both M$_{1}^{-}$ and M$_{2}^{-}$ reach maximum oscillator strength at doping density of one electron per moir\'e unit cell ($\nu=1$), and the decrease in the oscillator strength at electron densities of $\nu=1\pm \epsilon$ indicates the formation of an isolated Hubbard band at the incompressible state for $\nu=1$ \cite{Ciorciaro2023}.

The transfer of the oscillator strength from the charge-neutral moir\'e exciton M$_{1}$ to the M$_{2}^{-}$ exciton-polaron state is most obvious in the regime of coherent light-matter coupling as we add the structured top DBR mirror to form the microcavity. By sweeping the fundamental Laguerre-Gaussian mode through the resonances of M$_{1}$ and M$_{2}^{-}$, we observe clear anticrossings as hallmarks of strong light-matter coupling (see Fig.~\ref{fig1}d upper and lower panel). The respective polariton branches are captured by a coupled oscillator formalism, with Rabi splittings of $\Omega=9.0$~meV for M$_{1}$ at V$_{G}=0$~V and $\Omega=6.2$~meV for M$_{2}^{-}$ at V$_{G}=11.5$~V. The evolution of the Rabi splittings as a function of the electron doping is shown in Fig.~\ref{fig1}e for gate voltages where the strong coupling condition $\Omega>(\gamma+\kappa)/2$ is satisfied for both M$_1$ and M$_{2}^{-}$, with respective linewidths of $\gamma=7.6$ and $9.5$~meV. The homogeneous linewidth $\kappa=1.2$~meV of the cavity mode was determined from the Lorentzian part of a Voigt profile fit of the cavity lineshape (inhomogeneous contributions due to cavity vibrations were negligible on the time scale of coherent light-matter coupling; Supplementary Note~2). With increasing electron density, the Rabi splitting of M$_{1}$ decreased linearly and collapsed around $\nu>0.5$, while at the same time the anticrossing of the M$_{2}^{-}$ polariton reached its maximum value at electron filling $\nu=1$ before decreasing again at higher gate voltages.

\begin{figure}[t]  
\includegraphics[width=\columnwidth]{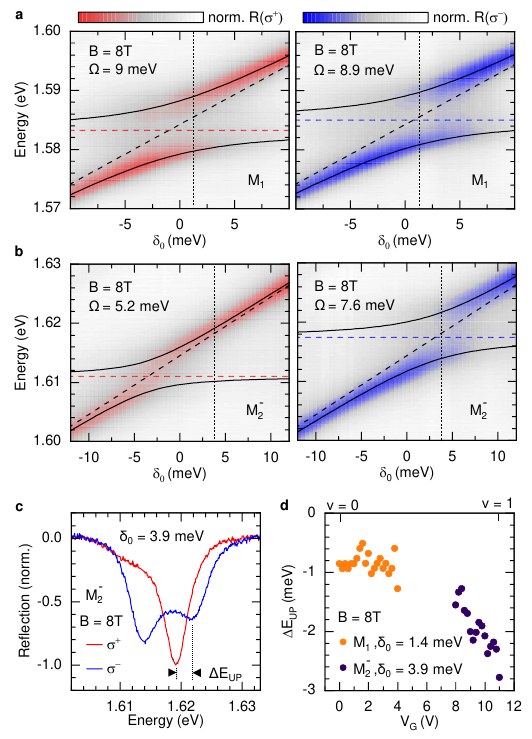}
\caption{\textbf{Magneto-optics of neutral and charged moir\'e polaritons at large magnetic field.} \textbf{a} and \textbf{b}, Polarization resolved reflection of M$_{1}$ exciton-polaritons and M$_{2}^{-}$ polaron-polaritons at V$_{G}=0$ and $11$~V, respectively, in a magnetic field of $B = 8$~T as a function of the cavity length $\delta _{\text{0}}$. Black solid lines show the fits of the coupled oscillator model for $\sigma^{+}$ and $\sigma^{-}$ circular polarization, and the energies of the corresponding bare excitons are shown by red and blue dashed lines; the field-independent cavity dispersion is shown by back dashed lines. \textbf{c}, Polarization-resolved spectra of M$_2^-$ polaron-polariton with the Zeeman splitting of the upper polariton branch $\Delta E_{\text{UP}}$, shown for a cavity length corresponding to the on-resonance condition for $\sigma^{-}$ polarization as indicated by dotted vertical lines in \textbf{b}. \textbf{d}, $\Delta E_{\text{UP}}$ of M$_{1}$ and M$_{2}^{-}$ as a function of V$_{G}$ in gate voltage regimes of strong coupling.}
\label{fig2}
\end{figure}

\begin{figure*}[t]  
\includegraphics[scale=1.0]{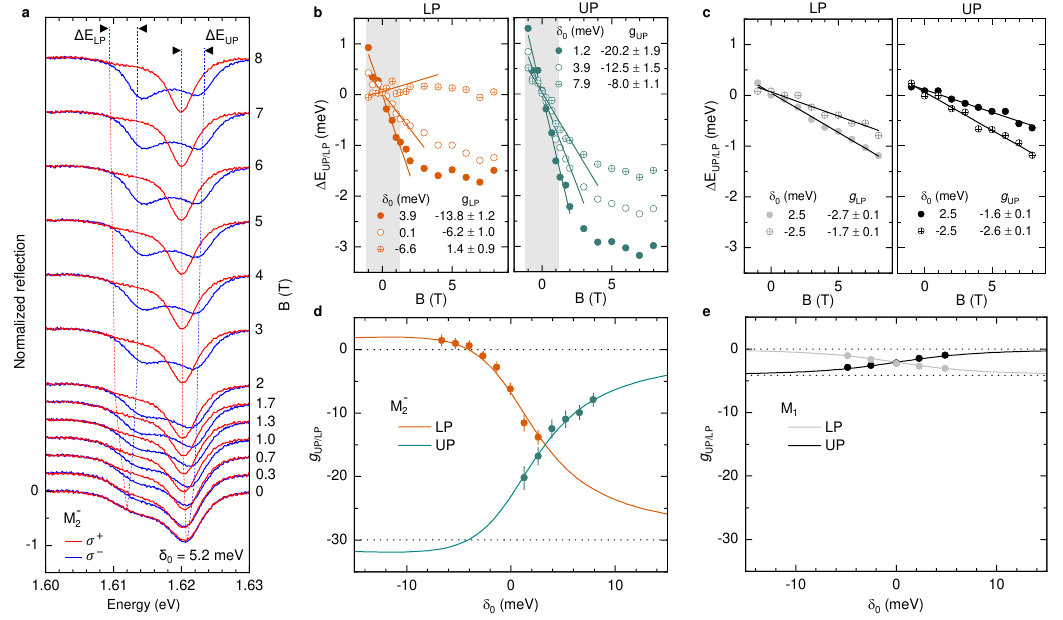}
\caption{\textbf{Magneto-optics of neutral and charged moir\'e polaritons at small magnetic fields.} \textbf{a}, Polarization-contrasting reflection spectra of M$_{2}^{-}$ polaron-polariton with increasing magnetic field for a fixed resonance detuning $\delta _{\text{0}}=5.2$~meV at $B = 0$~T. The dotted red and blue lines are guides to the eye. \textbf{b}, Nonlinear effective Zeeman splitting $\Delta E_{\text{LP/UP}}$ of M$_{2}^{-}$ lower (upper) polariton branch in the left (right) panel as a function of the magnetic field for selected detunings $\delta _{\text{0}}$. Solid lines show linear fits to the data at small fields ($|B|\leq1$~T highlighted in grey), yielding the effective upper/lower polariton $g$-factors $g_{\text{UP/LP}}$. \textbf{c}, Zeeman splittings $\Delta E_{\text{LP/UP}}$ of M$_1$ lower (upper) polariton branch in the left (right) panel exhibit linear dispersion over the whole range of magnetic fields for $\delta _{\text{0}}=\pm 2.5$~meV. \textbf{d}, Effective Land\'e factors $g_{\text{UP/LP}}$ for M$_{2}^{-}$ lower (orange) and upper (turquoise) polariton branch calculated from the coupled oscillator model for small magnetic fields $|B| \leq 1$~T; the data (circles in corresponding colors) are shown for various detunings $\delta _{\text{0}}$ with error bars from linear fits as in \textbf{b}. \textbf{e}, Calculated $g_{\text{UP/LP}}$ of the M$_1$ upper (black) and lower (grey) polariton branch with a symmetric dependence on $\delta _{\text{0}}$ and effective polariton $g$-factors (circles) obtained from linear fits as in \textbf{c}. The lower horizontal dashed lines in \textbf{d} and \textbf{e} indicate the $g$-factors of bare M$_{2}^{-}$ and M$_1$ states, respectively.} 
\label{fig3}
\end{figure*}

The most remarkable behavior of the strongly coupled exciton-polariton system in the presence of electron doping is induced by an external magnetic field. In polarization-contrasting reflection R($\sigma^{+/-}$) of M$_{1}$ and M$_{2}^{-}$ polaritons at $8$~T, shown in Fig.~\ref{fig2}a and b, we observe polarized polariton branches as a function of the cavity length which is represented in units of zero-field resonance detuning $\delta _{\text{0}}=E_C-E_X$ between the cavity and exciton energies $E_C$ and $E_X$ at $B = 0$~T. Upon cavity detuning, the $\sigma^{+}$ and $\sigma^{-}$ circularly polarized branches of the M$_{2}^{-}$ polaron-polariton in the left and right panels of Fig.~\ref{fig2}b exhibit avoided crossings with respective polarization-specific on-resonance Rabi splittings of $5.2$ and $7.5$~meV and an exciton Zeeman splitting (i. e. the energy difference between the red and blue dashed horizontal lines in Fig.~\ref{fig2}b) of $\sim 7$~meV. The latter contrasts the modest exciton Zeeman splitting extracted from the M$_{1}$ polariton in Fig.~\ref{fig2}a with a value of $2$~meV, comparable to the neutral exciton Zeeman splitting in MoSe$_2$ monolayers \cite{Li2014,MacNeill2015}.

With the cavity length set to $\delta _{\text{0}}=1.4$ (3.9) meV around the on-resonance condition for the $\sigma^-$ branch of the M$_{1}$ (M$_{2}^-$) polariton at $B = 8$~T, we determine the effective Zeeman splitting between the peak energies of the upper/lower polariton branch with $\sigma^+$ and $\sigma^-$ polarization as $\Delta E_{\text{UP/LP}} = E^{+}_{\text{UP/LP}} -E^{-}_{\text{UP/LP}}$, shown explicitly in Fig.~\ref{fig2}c for the upper polariton branch of M$_2^-$ at V$_G=11$~V. Remarkably, in this large-field limit, the evolution of the effective Zeeman splitting between the two circularly polarized upper polariton branches with the gate voltage in Fig.~\ref{fig2}d shows very distinct behaviors for M$_{1}$ and M$_{2}^-$ polaritons. As the voltage-induced electron filling per moir\'e cell is increased from $0$ to $1$, the upper polariton splitting $\Delta E_{\text{UP}}$ remains constant at $-1$~meV in the voltage range of M$_{1}$ between $0$ and $4$~V, then undergoes a jump to $-1.5$~meV with the onset of M$_{2}^-$ at $8$~V, and evolves gradually to its lowest value close to $-3$~meV at $11$~V. 
 
To understand this behavior, we first note that by setting the cavity resonance to zero spectral detuning for the bare exciton Zeeman branch with $\sigma^{-}$ polarization (crossing points of blue and black dashed lines in the right panels of Fig.~\ref{fig2}a and b), the cavity mode is blue-detuned by the exciton Zeeman splitting with respect to the $\sigma^{+}$ polarized exciton Zeeman branch. Since the presence of the M$_{2}^{-}$ exciton resonance is accompanied by an increase in the oscillator strength and thus in the Rabi splitting $\Omega$ up to electron filling of $\nu=1$ (Fig.~\ref{fig1}e), the respective on-resonance $\sigma^{-}$ polariton branch is affected by the gate voltage, whereas the $\sigma^{+}$ branch remains almost unchanged due to its primarily photonic nature (with photon Hopfield coefficient of $\sim 0.85$; Supplementary Note~3). Thus, the energy splitting $\Delta E_{\text{UP}}$ between the two circularly polarized upper polariton branches is the combined result of the exciton Zeeman splitting and the polarization-dependent Rabi splitting for a given cavity-exciton resonance detuning. Consequently, the constant and monotonously evolving Zeeman splittings $\Delta E_{\text{UP}}$ of the M$_1$ and M$_2^-$ upper polaritons in Fig.~\ref{fig2}d reflect the respective absence and presence of doping-induced magnetism, which peaks at electron doping of $\nu=1$ and thus yields the maximum Zeeman splitting for the M$_2^-$ upper polariton with a value roughly three times that of M$_{1}$. 

We point out that both $\sigma^{+}$ and $\sigma^{-}$ branches are clearly pronounced for both M$_1$ and M$_2^-$ polaritons even in the large-field limit (Fig.~\ref{fig2}a and b). Consequently, the enhancement in the polariton Zeeman splitting is distinct from the apparently giant paramagnetic response of MoSe$_2$ monolayer polaron \cite{Back_2017} and its polaron-polariton counterpart \cite{Lyons_2022}. There, at high magnetic fields, the respective $\sigma^{+}$ polarized lower-energy Zeeman exciton and polariton branches exhibit nearly complete quenching of the oscillator strength due to Pauli blocking, which leads to an effectively enhanced splitting \cite{Back_2017,Lyons_2022}. Here, field-dependent Rabi splitting is completely absent for charge-neutral M$_1$ exciton-polaritons, just like for the neutral exciton in monolayer MoSe$_2$ \cite{Koperski2019,Goryca2019}. Electron doping conditions a change in the oscillator strength for the polarized branches of the M$_{2}^{-}$ polaron-polariton with magnetic field, yielding different Rabi splittings for polariton branches with $\sigma^{+}$ and $\sigma^{-}$ polarization at $8$~T (Fig.~\ref{fig2}b). This moderate reduction of the oscillator strength as opposed to the complete quenching of one polarization in the case of the attractive polaron in MoSe$_2$ monolayers is the result of the spatial separation between the M$_{2}^{-}$ exciton wavefunction and the moir\'e lattice sites of ordered and spin-polarized electrons (see Supplementary Note~3).

In the following, we focus on electron-dressed moir\'e polaritons at low magnetic fields ($|B| \leq 1$~T), where correlated magnetism of charge-spin lattices is most pronounced \cite{Ciorciaro2023,Tang_2020,Xu2022,Campbell_2022,Tang_2023,Polovnikov2024}. For a fixed cavity detuning $\delta _{\text{0}}=5.2$~meV, the M$_{2}^{-}$ polaron-polariton branches develop pronounced Zeeman shifts already at small magnetic fields, as obvious from Fig.~\ref{fig3}a. Moreover, the Zeeman splittings of the upper and lower polariton branches $\Delta E_{\text{UP}}$ and $\Delta E_{\text{LP}}$, shown in Fig.~\ref{fig3}b for different cavity detunings $\delta _{\text{0}}$, follow a highly nonlinear evolution with enhanced effective Land\'e factors in the low-field limit. This behavior is analogous to the nonlinear evolution of moir\'e polaron $g$-factors in the presence of magnetic interactions on triangular electron-spin lattices \cite{Ciorciaro2023,Tang_2020,Xu2022,Campbell_2022,Tang_2023,Polovnikov2024}. The Zeeman splitting of the neutral M$_1$ exciton-polariton, in contrast, is linear throughout the whole range of magnetic fields in Fig.~\ref{fig3}c, and the corresponding effective polariton $g$-factor stems from the conventional paramagnetic response of TMD monolayer exciton-polaritons \cite{Lundt_2019}.

Consistently, the effective Land\'e factors of the M$_1$ upper and lower polariton branches are determined by the light and matter constituent according to the Hopfield coefficients, which in turn are controlled by the cavity resonance detuning. In the absence of doping, the effective polariton Land\'e factors derive from the linear Zeeman shift of the neutral moir\'e exciton as shown in Fig.~\ref{fig3}e. The symmetric evolution of $g_{\text{UP/LP}}$ around $\delta _{\text{0}}=0$ is asymptotically bounded at large cavity detunings by the photon and exciton $g$-factor of $0$ and $-4$, respectively. Remarkably, this symmetry is absent for M$_{2}^{-}$ polaron-polaritons in Fig.~\ref{fig3}d: While $g_{\text{UP}}$ exhibits a large negative value at a positive detuning $\delta _{\text{0}}=7.9$~meV ($g_{\text{UP}}=-8.0$ in the right panel of Fig.~\ref{fig3}b), its $g_{\text{LP}}$ counterpart is much smaller at the corresponding negative detuning $\delta _{\text{0}}=-6.6$~meV ($g_{\text{LP}}=1.4$ in the left panel of Fig.~\ref{fig3}b) and, strikingly, even turns positive. 

This feature is inexplicable on the basis of the underlying bare exciton $g$-factor alone. In fact, it is a prime consequence of correlated magnetism on the electron-spin lattice, which conditions polarization-sensitive Rabi splitting of the M$_{2}^{-}$ polaron-polariton at finite magnetic fields. Taking the linear dependence of the polarization-contrasting Rabi splitting at small magnetic fields into account (Supplementary Note~3), our model captures the overall evolution of M$_{2}^{-}$ effective polariton Land\'e factors $g_{\text{UP/LP}}$ with cavity detuning in Fig.~\ref{fig3}d, and in particular the sign-reversal of the lower polariton $g$-factor for negative detunings. Remarkably, the model also predicts for large negative detunings an enhancement of the upper polariton Land\'e factor beyond the $g$-factor of the moir\'e polaron counterpart.

In conclusion, our work establishes moir\'e exciton-polaritons as a novel regime of quantum many-body phenomena on triangular exciton, electron and spin lattices. Dressed by cavity photons and spatially ordered electrons, moir\'e excitons form polaron-polariton states with signatures of correlated magnetism that manifest as highly nonlinear effective polariton Land\'e factors around unity electron filling per moir\'e cell. Intriguingly, imbalancing the polariton fractions by cavity-resonance detuning allows to control the effective polaron-polariton $g$-factor continuously, even to values below the bare exciton $g$-factor and above zero. Our demonstration of correlated magnetism for moir\'e polaron-polaritons is a crucial step towards utilizing moir\'e lattices and strong-light matter coupling to control condensed matter quantum many-body systems. It also paves the way towards other correlated polaritonic phases of ferromagnetic \cite{Ciorciaro2023,Ma_2022}, topological \cite{Liu_2022} or fractional quantum Hall states \cite{Ravets_2018,Knueppel_2019} and quantum condensates \cite{Leggett2006} of strongly coupled electron-exciton gases. 

\vspace{17pt}
\noindent \textbf{Methods}\\
\noindent \textbf{Device fabrication:} TMD monolayers were synthesized by chemical vapor deposition and assembled with few-layer graphite and hBN flakes via the dry transfer method. The TMD heterostructure was placed on top of a DBR made of $7$ pairs of SiO$_{2}$/TiO$_{2}$ layers, with the terminating layer of TiO$_{2}$ forming the bottom part of the optical microcavity at a field node. To maximize the intensity of the reflected light at the position of the TMD heterostack inside the optical cavity for the relevant wavelength range, the thickness of the bottom hBN flake was matched to the effective optical path of a quarter wavelength of the lowest-energy moir\'e exciton transition. The thickness of the top and bottom hBN flakes was chosen identical to ensure a symmetric field-effect device in the dual-gate configuration. he top part of the cavity was built from a glass mesa containing concave spherical cap indentations with $300$~nm depth and a diameter of $6~\mu$m formed by gallium focused ion beam milling and coated with $5$ pairs of TiO$_{2}$/SiO$_{2}$ layers with SiO$_{2}$ terminating layer. The photonic trapping potential resulting from the spherical indentation supports stable Laguerre-Gaussian modes at small cavity lengths.
\vspace{6pt} 

\noindent \textbf{Optical spectroscopy:} Cryogenic DR spectroscopy in absence of the top DBR was conducted using a home-built confocal microscope in back-scattering geometry. The sample was loaded in a closed-cycle-cryostat (attocube systems, attoDRY1000) with a base temperature of $3.2$~K and a superconducting magnet providing magnetic fields of up to $\pm 9$~T. Piezo-stepping and scanning units (attocube systems) were used for sample positioning with respect to a low-temperature apochromatic objective (attocube systems). A supercontinuum laser (NKT Photonics) was used as broadband light source, the reflection signal was spectrally dispersed by a monochromator (Roper Scientific, Acton SP2500 with 300 grooves/mm grating) and detected by a liquid nitrogen cooled CCD (Roper Scientific, Spec-10:100BR). The DR spectra were obtained by normalizing the reflection spectra ($R$) from the HBL region by the spectrum from the sample region without HBL ($R_0$) as $\textrm{DR} = (R-R_0)/R_0$. A custom-modified low vibration closed-cycle cryostat (attocube systems, attoDRY1000) was used for the microcavity experiments. A broadband quartz tungsten-halogen lamp was focused through the top mirror into the cavity by a lens (Thorlabs, 354105-B) to form a Gaussian spot size with a full-width at half-maximum diameter of $< 5.0 ~\mu$m. The physical distance between the microcavity mirrors was controlled via nano-positioners with nm accuracy. All cavity measurements were performed using the fundamental longitudinal cavity mode of order $\text{q} = 4$, the lowest possible mode order without introducing physical contact between top and bottom DBR mirrors, and employing a spectrometer (Andor Shamrock 500i) equipped with a CCD (Andor iKon-M 934).

\noindent \textbf{Acknowledgements}\\
\noindent This research was funded by the European Research Council (ERC) under the Grant Agreement No.~772195, as well as the Deutsche Forschungsgemeinschaft (DFG, German Research Foundation) within the Priority Programme SPP~2244 2DMP, the Germany's Excellence Strategy under grant No.~EXC-2111-390814868 (MCQST), and the projects SCHN1376 11-1 and SCHN1376 14-1. Z.\,L. was supported by the China Scholarship Council grant No.~201808140196, A.\,B. by the European Union's Framework Programme for Research and Innovation Horizon 2020 (2014--2020) under the Marie Sk{\l}odowska-Curie grant agreement No.~754388 (LMUResearchFellows) and LMUexcellent, funded by the Federal Ministry of Education and Research (BMBF) and the Free State of Bavaria under the Excellence Strategy of the German Federal Government and the L{\"a}nder. L.\,H. and A.\,H. acknowledge funding by the Bavarian Hightech Agenda project EQAP. K.\,W. and T.\,T. acknowledge support from the JSPS KAKENHI (Grant No. 20H00354 and 23H02052) and World Premier International Research Center Initiative (WPI), MEXT, Japan.
\vspace{6pt}
\\
\noindent \textbf{Contributions:} J.\,G. and Z.\,L. synthesized monolayers by vapor deposition. J.\,S. designed and fabricated the van der Waals heterostructure. J.\,S., B.\,P. and L.\,H. performed confocal spectroscopy and cavity experiments at the LMU Munich. L.\,L., B.\,H. realized the magneto-optical cavity. L.\,L., J.\,S., B.\,H. and J.-C.\,D. performed the experiments at the University of Oldenburg. M.\,E, C.\,B., and F.\,E. implemented structured DBRs. K.\,W. and T.\,T. provided high quality hBN. J.\,S. and A.\,S.\,B.  developed the theoretical model. J.\,S., L.\,L., A.\,S.\,B., C.\,S. and A.\,H. analyzed and interpreted the data and wrote the manuscript with input from all coauthors. J.\,S. and L.\,L. contributed equally to this work.
\vspace{6pt}
\\
\noindent \textbf{Corresponding authors:} C.\,S. (christian.schneider@uni-oldenburg.de) and A.\,H. (alexander.hoegele@lmu.de)
\vspace{6pt}
\\
\noindent \textbf{Data availability:} Source data are provided with this manuscript. Additional data that support the findings of this study are available from the corresponding authors upon reasonable request.
\vspace{6pt}
\\
\noindent\textbf{Code availability:} The codes that support the findings of this study are available from the corresponding authors upon reasonable request.
\vspace{6pt}
\\
\noindent\textbf{Competing interests:} The authors declare no competing interests.


\begin{thebibliography}{52}%
\makeatletter
\providecommand \@ifxundefined [1]{%
 \@ifx{#1\undefined}
}%
\providecommand \@ifnum [1]{%
 \ifnum #1\expandafter \@firstoftwo
 \else \expandafter \@secondoftwo
 \fi
}%
\providecommand \@ifx [1]{%
 \ifx #1\expandafter \@firstoftwo
 \else \expandafter \@secondoftwo
 \fi
}%
\providecommand \natexlab [1]{#1}%
\providecommand \enquote  [1]{``#1''}%
\providecommand \bibnamefont  [1]{#1}%
\providecommand \bibfnamefont [1]{#1}%
\providecommand \citenamefont [1]{#1}%
\providecommand \href@noop [0]{\@secondoftwo}%
\providecommand \href [0]{\begingroup \@sanitize@url \@href}%
\providecommand \@href[1]{\@@startlink{#1}\@@href}%
\providecommand \@@href[1]{\endgroup#1\@@endlink}%
\providecommand \@sanitize@url [0]{\catcode `\\12\catcode `\$12\catcode
  `\&12\catcode `\#12\catcode `\^12\catcode `\_12\catcode `\%12\relax}%
\providecommand \@@startlink[1]{}%
\providecommand \@@endlink[0]{}%
\providecommand \url  [0]{\begingroup\@sanitize@url \@url }%
\providecommand \@url [1]{\endgroup\@href {#1}{\urlprefix }}%
\providecommand \urlprefix  [0]{URL }%
\providecommand \Eprint [0]{\href }%
\providecommand \doibase [0]{https://doi.org/}%
\providecommand \selectlanguage [0]{\@gobble}%
\providecommand \bibinfo  [0]{\@secondoftwo}%
\providecommand \bibfield  [0]{\@secondoftwo}%
\providecommand \translation [1]{[#1]}%
\providecommand \BibitemOpen [0]{}%
\providecommand \bibitemStop [0]{}%
\providecommand \bibitemNoStop [0]{.\EOS\space}%
\providecommand \EOS [0]{\spacefactor3000\relax}%
\providecommand \BibitemShut  [1]{\csname bibitem#1\endcsname}%
\let\auto@bib@innerbib\@empty
\bibitem [{\citenamefont {Wu}\ \emph {et~al.}(2018{\natexlab{a}})\citenamefont
  {Wu}, \citenamefont {Lovorn}, \citenamefont {Tutuc},\ and\ \citenamefont
  {MacDonald}}]{WuHubbard2018}%
  \BibitemOpen
  \bibfield  {author} {\bibinfo {author} {\bibfnamefont {F.}~\bibnamefont
  {Wu}}, \bibinfo {author} {\bibfnamefont {T.}~\bibnamefont {Lovorn}}, \bibinfo
  {author} {\bibfnamefont {E.}~\bibnamefont {Tutuc}},\ and\ \bibinfo {author}
  {\bibfnamefont {A.~H.}\ \bibnamefont {MacDonald}},\ }\bibfield  {title}
  {\bibinfo {title} {Hubbard model physics in transition metal dichalcogenide
  moir\'e bands},\ }\href {https://doi.org/10.1103/PhysRevLett.121.026402}
  {\bibfield  {journal} {\bibinfo  {journal} {Phys. Rev. Lett.}\ }\textbf
  {\bibinfo {volume} {121}},\ \bibinfo {pages} {026402} (\bibinfo {year}
  {2018}{\natexlab{a}})}\BibitemShut {NoStop}%
\bibitem [{\citenamefont {Polovnikov}\ \emph
  {et~al.}(2024{\natexlab{a}})\citenamefont {Polovnikov}, \citenamefont
  {Scherzer}, \citenamefont {Misra}, \citenamefont {Schl\"omer}, \citenamefont
  {Trapp}, \citenamefont {Huang}, \citenamefont {Mohl}, \citenamefont {Li},
  \citenamefont {G\"oser}, \citenamefont {F\"orste}, \citenamefont {Bilgin},
  \citenamefont {Watanabe}, \citenamefont {Taniguchi}, \citenamefont {Bohrdt},
  \citenamefont {Grusdt}, \citenamefont {Baimuratov},\ and\ \citenamefont
  {H\"ogele}}]{Polovnikov2024}%
  \BibitemOpen
  \bibfield  {author} {\bibinfo {author} {\bibfnamefont {B.}~\bibnamefont
  {Polovnikov}}, \bibinfo {author} {\bibfnamefont {J.}~\bibnamefont
  {Scherzer}}, \bibinfo {author} {\bibfnamefont {S.}~\bibnamefont {Misra}},
  \bibinfo {author} {\bibfnamefont {H.}~\bibnamefont {Schl\"omer}}, \bibinfo
  {author} {\bibfnamefont {J.}~\bibnamefont {Trapp}}, \bibinfo {author}
  {\bibfnamefont {X.}~\bibnamefont {Huang}}, \bibinfo {author} {\bibfnamefont
  {C.}~\bibnamefont {Mohl}}, \bibinfo {author} {\bibfnamefont {Z.}~\bibnamefont
  {Li}}, \bibinfo {author} {\bibfnamefont {J.}~\bibnamefont {G\"oser}},
  \bibinfo {author} {\bibfnamefont {J.}~\bibnamefont {F\"orste}}, \bibinfo
  {author} {\bibfnamefont {I.}~\bibnamefont {Bilgin}}, \bibinfo {author}
  {\bibfnamefont {K.}~\bibnamefont {Watanabe}}, \bibinfo {author}
  {\bibfnamefont {T.}~\bibnamefont {Taniguchi}}, \bibinfo {author}
  {\bibfnamefont {A.}~\bibnamefont {Bohrdt}}, \bibinfo {author} {\bibfnamefont
  {F.}~\bibnamefont {Grusdt}}, \bibinfo {author} {\bibfnamefont {A.~S.}\
  \bibnamefont {Baimuratov}},\ and\ \bibinfo {author} {\bibfnamefont
  {A.}~\bibnamefont {H\"ogele}},\ }\bibfield  {title} {\bibinfo {title}
  {{Implementation of the bilayer Hubbard model in a moir\'e
  heterostructure}},\ }\href@noop {} {\bibfield  {journal} {\bibinfo  {journal}
  {arXiv preprint arXiv:2404.05494}\ } (\bibinfo {year}
  {2024}{\natexlab{a}})}\BibitemShut {NoStop}%
\bibitem [{\citenamefont {Xu}\ \emph {et~al.}(2020)\citenamefont {Xu},
  \citenamefont {Liu}, \citenamefont {Rhodes}, \citenamefont {Watanabe},
  \citenamefont {Taniguchi}, \citenamefont {Hone}, \citenamefont {Elser},
  \citenamefont {Mak},\ and\ \citenamefont {Shan}}]{Xu2020}%
  \BibitemOpen
  \bibfield  {author} {\bibinfo {author} {\bibfnamefont {Y.}~\bibnamefont
  {Xu}}, \bibinfo {author} {\bibfnamefont {S.}~\bibnamefont {Liu}}, \bibinfo
  {author} {\bibfnamefont {D.~A.}\ \bibnamefont {Rhodes}}, \bibinfo {author}
  {\bibfnamefont {K.}~\bibnamefont {Watanabe}}, \bibinfo {author}
  {\bibfnamefont {T.}~\bibnamefont {Taniguchi}}, \bibinfo {author}
  {\bibfnamefont {J.}~\bibnamefont {Hone}}, \bibinfo {author} {\bibfnamefont
  {V.}~\bibnamefont {Elser}}, \bibinfo {author} {\bibfnamefont {K.~F.}\
  \bibnamefont {Mak}},\ and\ \bibinfo {author} {\bibfnamefont {J.}~\bibnamefont
  {Shan}},\ }\bibfield  {title} {\bibinfo {title} {Correlated insulating states
  at fractional fillings of moiré superlattices},\ }\href
  {https://doi.org/10.1038/s41586-020-2868-6} {\bibfield  {journal} {\bibinfo
  {journal} {Nature}\ }\textbf {\bibinfo {volume} {587}},\ \bibinfo {pages}
  {214} (\bibinfo {year} {2020})}\BibitemShut {NoStop}%
\bibitem [{\citenamefont {Regan}\ \emph {et~al.}(2020)\citenamefont {Regan},
  \citenamefont {Wang}, \citenamefont {Jin}, \citenamefont {Bakti~Utama},
  \citenamefont {Gao}, \citenamefont {Wei}, \citenamefont {Zhao}, \citenamefont
  {Zhao}, \citenamefont {Zhang}, \citenamefont {Yumigeta}, \citenamefont
  {Blei}, \citenamefont {Carlström}, \citenamefont {Watanabe}, \citenamefont
  {Taniguchi}, \citenamefont {Tongay}, \citenamefont {Crommie}, \citenamefont
  {Zettl},\ and\ \citenamefont {Wang}}]{Regan2020}%
  \BibitemOpen
  \bibfield  {author} {\bibinfo {author} {\bibfnamefont {E.~C.}\ \bibnamefont
  {Regan}}, \bibinfo {author} {\bibfnamefont {D.}~\bibnamefont {Wang}},
  \bibinfo {author} {\bibfnamefont {C.}~\bibnamefont {Jin}}, \bibinfo {author}
  {\bibfnamefont {M.~I.}\ \bibnamefont {Bakti~Utama}}, \bibinfo {author}
  {\bibfnamefont {B.}~\bibnamefont {Gao}}, \bibinfo {author} {\bibfnamefont
  {X.}~\bibnamefont {Wei}}, \bibinfo {author} {\bibfnamefont {S.}~\bibnamefont
  {Zhao}}, \bibinfo {author} {\bibfnamefont {W.}~\bibnamefont {Zhao}}, \bibinfo
  {author} {\bibfnamefont {Z.}~\bibnamefont {Zhang}}, \bibinfo {author}
  {\bibfnamefont {K.}~\bibnamefont {Yumigeta}}, \bibinfo {author}
  {\bibfnamefont {M.}~\bibnamefont {Blei}}, \bibinfo {author} {\bibfnamefont
  {J.~D.}\ \bibnamefont {Carlström}}, \bibinfo {author} {\bibfnamefont
  {K.}~\bibnamefont {Watanabe}}, \bibinfo {author} {\bibfnamefont
  {T.}~\bibnamefont {Taniguchi}}, \bibinfo {author} {\bibfnamefont
  {S.}~\bibnamefont {Tongay}}, \bibinfo {author} {\bibfnamefont
  {M.}~\bibnamefont {Crommie}}, \bibinfo {author} {\bibfnamefont
  {A.}~\bibnamefont {Zettl}},\ and\ \bibinfo {author} {\bibfnamefont
  {F.}~\bibnamefont {Wang}},\ }\bibfield  {title} {\bibinfo {title} {{Mott and
  generalized {W}igner crystal states in {WSe}$_2$/{WS}$_2$ moir\'e
  superlattices}},\ }\href {https://doi.org/10.1038/s41586-020-2092-4}
  {\bibfield  {journal} {\bibinfo  {journal} {Nature}\ }\textbf {\bibinfo
  {volume} {579}},\ \bibinfo {pages} {359} (\bibinfo {year}
  {2020})}\BibitemShut {NoStop}%
\bibitem [{\citenamefont {Shimazaki}\ \emph {et~al.}(2020)\citenamefont
  {Shimazaki}, \citenamefont {Schwartz}, \citenamefont {Watanabe},
  \citenamefont {Taniguchi}, \citenamefont {Kroner},\ and\ \citenamefont
  {Imamoğlu}}]{Shimazaki2020}%
  \BibitemOpen
  \bibfield  {author} {\bibinfo {author} {\bibfnamefont {Y.}~\bibnamefont
  {Shimazaki}}, \bibinfo {author} {\bibfnamefont {I.}~\bibnamefont {Schwartz}},
  \bibinfo {author} {\bibfnamefont {K.}~\bibnamefont {Watanabe}}, \bibinfo
  {author} {\bibfnamefont {T.}~\bibnamefont {Taniguchi}}, \bibinfo {author}
  {\bibfnamefont {M.}~\bibnamefont {Kroner}},\ and\ \bibinfo {author}
  {\bibfnamefont {A.}~\bibnamefont {Imamoğlu}},\ }\bibfield  {title} {\bibinfo
  {title} {Strongly correlated electrons and hybrid excitons in a moir\'e
  heterostructure},\ }\href {https://doi.org/10.1038/s41586-020-2191-2}
  {\bibfield  {journal} {\bibinfo  {journal} {Nature}\ }\textbf {\bibinfo
  {volume} {580}},\ \bibinfo {pages} {472} (\bibinfo {year}
  {2020})}\BibitemShut {NoStop}%
\bibitem [{\citenamefont {Serlin}\ \emph {et~al.}(2020)\citenamefont {Serlin},
  \citenamefont {Tschirhart}, \citenamefont {Polshyn}, \citenamefont {Zhang},
  \citenamefont {Zhu}, \citenamefont {Watanabe}, \citenamefont {Taniguchi},
  \citenamefont {Balents},\ and\ \citenamefont {Young}}]{Serlin2020}%
  \BibitemOpen
  \bibfield  {author} {\bibinfo {author} {\bibfnamefont {M.}~\bibnamefont
  {Serlin}}, \bibinfo {author} {\bibfnamefont {C.~L.}\ \bibnamefont
  {Tschirhart}}, \bibinfo {author} {\bibfnamefont {H.}~\bibnamefont {Polshyn}},
  \bibinfo {author} {\bibfnamefont {Y.}~\bibnamefont {Zhang}}, \bibinfo
  {author} {\bibfnamefont {J.}~\bibnamefont {Zhu}}, \bibinfo {author}
  {\bibfnamefont {K.}~\bibnamefont {Watanabe}}, \bibinfo {author}
  {\bibfnamefont {T.}~\bibnamefont {Taniguchi}}, \bibinfo {author}
  {\bibfnamefont {L.}~\bibnamefont {Balents}},\ and\ \bibinfo {author}
  {\bibfnamefont {A.~F.}\ \bibnamefont {Young}},\ }\bibfield  {title} {\bibinfo
  {title} {{Intrinsic quantized anomalous Hall effect in a moir\'e
  heterostructure}},\ }\href {https://doi.org/10.1126/science.aay5533}
  {\bibfield  {journal} {\bibinfo  {journal} {Science}\ }\textbf {\bibinfo
  {volume} {367}},\ \bibinfo {pages} {900} (\bibinfo {year}
  {2020})}\BibitemShut {NoStop}%
\bibitem [{\citenamefont {Li}\ \emph {et~al.}(2021)\citenamefont {Li},
  \citenamefont {Jiang}, \citenamefont {Shen}, \citenamefont {Zhang},
  \citenamefont {Li}, \citenamefont {Tao}, \citenamefont {Devakul},
  \citenamefont {Watanabe}, \citenamefont {Taniguchi}, \citenamefont {Fu},
  \citenamefont {Shan},\ and\ \citenamefont {Mak}}]{LiMak2021}%
  \BibitemOpen
  \bibfield  {author} {\bibinfo {author} {\bibfnamefont {T.}~\bibnamefont
  {Li}}, \bibinfo {author} {\bibfnamefont {S.}~\bibnamefont {Jiang}}, \bibinfo
  {author} {\bibfnamefont {B.}~\bibnamefont {Shen}}, \bibinfo {author}
  {\bibfnamefont {Y.}~\bibnamefont {Zhang}}, \bibinfo {author} {\bibfnamefont
  {L.}~\bibnamefont {Li}}, \bibinfo {author} {\bibfnamefont {Z.}~\bibnamefont
  {Tao}}, \bibinfo {author} {\bibfnamefont {T.}~\bibnamefont {Devakul}},
  \bibinfo {author} {\bibfnamefont {K.}~\bibnamefont {Watanabe}}, \bibinfo
  {author} {\bibfnamefont {T.}~\bibnamefont {Taniguchi}}, \bibinfo {author}
  {\bibfnamefont {L.}~\bibnamefont {Fu}}, \bibinfo {author} {\bibfnamefont
  {J.}~\bibnamefont {Shan}},\ and\ \bibinfo {author} {\bibfnamefont {K.~F.}\
  \bibnamefont {Mak}},\ }\bibfield  {title} {\bibinfo {title} {{Quantum
  anomalous Hall effect from intertwined moir\'e bands}},\ }\href
  {https://doi.org/10.1038/s41586-021-04171-1} {\bibfield  {journal} {\bibinfo
  {journal} {Nature}\ }\textbf {\bibinfo {volume} {600}},\ \bibinfo {pages}
  {641} (\bibinfo {year} {2021})}\BibitemShut {NoStop}%
\bibitem [{\citenamefont {Cai}\ \emph {et~al.}(2023)\citenamefont {Cai},
  \citenamefont {Anderson}, \citenamefont {Wang}, \citenamefont {Zhang},
  \citenamefont {Liu}, \citenamefont {Holtzmann}, \citenamefont {Zhang},
  \citenamefont {Fan}, \citenamefont {Taniguchi}, \citenamefont {Watanabe},
  \citenamefont {Ran}, \citenamefont {Cao}, \citenamefont {Fu}, \citenamefont
  {Xiao}, \citenamefont {Yao},\ and\ \citenamefont {Xu}}]{Cai2023}%
  \BibitemOpen
  \bibfield  {author} {\bibinfo {author} {\bibfnamefont {J.}~\bibnamefont
  {Cai}}, \bibinfo {author} {\bibfnamefont {E.}~\bibnamefont {Anderson}},
  \bibinfo {author} {\bibfnamefont {C.}~\bibnamefont {Wang}}, \bibinfo {author}
  {\bibfnamefont {X.}~\bibnamefont {Zhang}}, \bibinfo {author} {\bibfnamefont
  {X.}~\bibnamefont {Liu}}, \bibinfo {author} {\bibfnamefont {W.}~\bibnamefont
  {Holtzmann}}, \bibinfo {author} {\bibfnamefont {Y.}~\bibnamefont {Zhang}},
  \bibinfo {author} {\bibfnamefont {F.}~\bibnamefont {Fan}}, \bibinfo {author}
  {\bibfnamefont {T.}~\bibnamefont {Taniguchi}}, \bibinfo {author}
  {\bibfnamefont {K.}~\bibnamefont {Watanabe}}, \bibinfo {author}
  {\bibfnamefont {Y.}~\bibnamefont {Ran}}, \bibinfo {author} {\bibfnamefont
  {T.}~\bibnamefont {Cao}}, \bibinfo {author} {\bibfnamefont {L.}~\bibnamefont
  {Fu}}, \bibinfo {author} {\bibfnamefont {D.}~\bibnamefont {Xiao}}, \bibinfo
  {author} {\bibfnamefont {W.}~\bibnamefont {Yao}},\ and\ \bibinfo {author}
  {\bibfnamefont {X.}~\bibnamefont {Xu}},\ }\bibfield  {title} {\bibinfo
  {title} {{Signatures of fractional quantum anomalous Hall states in twisted
  MoTe$_2$}},\ }\href {https://doi.org/10.1038/s41586-023-06289-w} {\bibfield
  {journal} {\bibinfo  {journal} {Nature}\ }\textbf {\bibinfo {volume} {622}},\
  \bibinfo {pages} {63} (\bibinfo {year} {2023})}\BibitemShut {NoStop}%
\bibitem [{\citenamefont {Park}\ \emph {et~al.}(2023)\citenamefont {Park},
  \citenamefont {Cai}, \citenamefont {Anderson}, \citenamefont {Zhang},
  \citenamefont {Zhu}, \citenamefont {Liu}, \citenamefont {Wang}, \citenamefont
  {Holtzmann}, \citenamefont {Hu}, \citenamefont {Liu}, \citenamefont
  {Taniguchi}, \citenamefont {Watanabe}, \citenamefont {Chu}, \citenamefont
  {Cao}, \citenamefont {Fu}, \citenamefont {Yao}, \citenamefont {Chang},
  \citenamefont {Cobden}, \citenamefont {Xiao},\ and\ \citenamefont
  {Xu}}]{Park2023}%
  \BibitemOpen
  \bibfield  {author} {\bibinfo {author} {\bibfnamefont {H.}~\bibnamefont
  {Park}}, \bibinfo {author} {\bibfnamefont {J.}~\bibnamefont {Cai}}, \bibinfo
  {author} {\bibfnamefont {E.}~\bibnamefont {Anderson}}, \bibinfo {author}
  {\bibfnamefont {Y.}~\bibnamefont {Zhang}}, \bibinfo {author} {\bibfnamefont
  {J.}~\bibnamefont {Zhu}}, \bibinfo {author} {\bibfnamefont {X.}~\bibnamefont
  {Liu}}, \bibinfo {author} {\bibfnamefont {C.}~\bibnamefont {Wang}}, \bibinfo
  {author} {\bibfnamefont {W.}~\bibnamefont {Holtzmann}}, \bibinfo {author}
  {\bibfnamefont {C.}~\bibnamefont {Hu}}, \bibinfo {author} {\bibfnamefont
  {Z.}~\bibnamefont {Liu}}, \bibinfo {author} {\bibfnamefont {T.}~\bibnamefont
  {Taniguchi}}, \bibinfo {author} {\bibfnamefont {K.}~\bibnamefont {Watanabe}},
  \bibinfo {author} {\bibfnamefont {J.-H.}\ \bibnamefont {Chu}}, \bibinfo
  {author} {\bibfnamefont {T.}~\bibnamefont {Cao}}, \bibinfo {author}
  {\bibfnamefont {L.}~\bibnamefont {Fu}}, \bibinfo {author} {\bibfnamefont
  {W.}~\bibnamefont {Yao}}, \bibinfo {author} {\bibfnamefont {C.-Z.}\
  \bibnamefont {Chang}}, \bibinfo {author} {\bibfnamefont {D.}~\bibnamefont
  {Cobden}}, \bibinfo {author} {\bibfnamefont {D.}~\bibnamefont {Xiao}},\ and\
  \bibinfo {author} {\bibfnamefont {X.}~\bibnamefont {Xu}},\ }\bibfield
  {title} {\bibinfo {title} {{Observation of fractionally quantized anomalous
  Hall effect}},\ }\href {https://doi.org/10.1038/s41586-023-06536-0}
  {\bibfield  {journal} {\bibinfo  {journal} {Nature}\ }\textbf {\bibinfo
  {volume} {622}},\ \bibinfo {pages} {74} (\bibinfo {year} {2023})}\BibitemShut
  {NoStop}%
\bibitem [{\citenamefont {Xu}\ \emph {et~al.}(2023)\citenamefont {Xu},
  \citenamefont {Sun}, \citenamefont {Jia}, \citenamefont {Liu}, \citenamefont
  {Xu}, \citenamefont {Li}, \citenamefont {Gu}, \citenamefont {Watanabe},
  \citenamefont {Taniguchi}, \citenamefont {Tong}, \citenamefont {Jia},
  \citenamefont {Shi}, \citenamefont {Jiang}, \citenamefont {Zhang},
  \citenamefont {Liu},\ and\ \citenamefont {Li}}]{Xu2023}%
  \BibitemOpen
  \bibfield  {author} {\bibinfo {author} {\bibfnamefont {F.}~\bibnamefont
  {Xu}}, \bibinfo {author} {\bibfnamefont {Z.}~\bibnamefont {Sun}}, \bibinfo
  {author} {\bibfnamefont {T.}~\bibnamefont {Jia}}, \bibinfo {author}
  {\bibfnamefont {C.}~\bibnamefont {Liu}}, \bibinfo {author} {\bibfnamefont
  {C.}~\bibnamefont {Xu}}, \bibinfo {author} {\bibfnamefont {C.}~\bibnamefont
  {Li}}, \bibinfo {author} {\bibfnamefont {Y.}~\bibnamefont {Gu}}, \bibinfo
  {author} {\bibfnamefont {K.}~\bibnamefont {Watanabe}}, \bibinfo {author}
  {\bibfnamefont {T.}~\bibnamefont {Taniguchi}}, \bibinfo {author}
  {\bibfnamefont {B.}~\bibnamefont {Tong}}, \bibinfo {author} {\bibfnamefont
  {J.}~\bibnamefont {Jia}}, \bibinfo {author} {\bibfnamefont {Z.}~\bibnamefont
  {Shi}}, \bibinfo {author} {\bibfnamefont {S.}~\bibnamefont {Jiang}}, \bibinfo
  {author} {\bibfnamefont {Y.}~\bibnamefont {Zhang}}, \bibinfo {author}
  {\bibfnamefont {X.}~\bibnamefont {Liu}},\ and\ \bibinfo {author}
  {\bibfnamefont {T.}~\bibnamefont {Li}},\ }\bibfield  {title} {\bibinfo
  {title} {{Observation of Integer and Fractional Quantum Anomalous Hall
  Effects in Twisted Bilayer ${\mathrm{MoTe}}_{2}$}},\ }\href
  {https://doi.org/10.1103/PhysRevX.13.031037} {\bibfield  {journal} {\bibinfo
  {journal} {Phys. Rev. X}\ }\textbf {\bibinfo {volume} {13}},\ \bibinfo
  {pages} {031037} (\bibinfo {year} {2023})}\BibitemShut {NoStop}%
\bibitem [{\citenamefont {Ciorciaro}\ \emph {et~al.}(2023)\citenamefont
  {Ciorciaro}, \citenamefont {Smoleński}, \citenamefont {Morera},
  \citenamefont {Kiper}, \citenamefont {Hiestand}, \citenamefont {Kroner},
  \citenamefont {Zhang}, \citenamefont {Watanabe}, \citenamefont {Taniguchi},
  \citenamefont {Demler},\ and\ \citenamefont {İmamoğlu}}]{Ciorciaro2023}%
  \BibitemOpen
  \bibfield  {author} {\bibinfo {author} {\bibfnamefont {L.}~\bibnamefont
  {Ciorciaro}}, \bibinfo {author} {\bibfnamefont {T.}~\bibnamefont
  {Smoleński}}, \bibinfo {author} {\bibfnamefont {I.}~\bibnamefont {Morera}},
  \bibinfo {author} {\bibfnamefont {N.}~\bibnamefont {Kiper}}, \bibinfo
  {author} {\bibfnamefont {S.}~\bibnamefont {Hiestand}}, \bibinfo {author}
  {\bibfnamefont {M.}~\bibnamefont {Kroner}}, \bibinfo {author} {\bibfnamefont
  {Y.}~\bibnamefont {Zhang}}, \bibinfo {author} {\bibfnamefont
  {K.}~\bibnamefont {Watanabe}}, \bibinfo {author} {\bibfnamefont
  {T.}~\bibnamefont {Taniguchi}}, \bibinfo {author} {\bibfnamefont
  {E.}~\bibnamefont {Demler}},\ and\ \bibinfo {author} {\bibfnamefont
  {A.}~\bibnamefont {İmamoğlu}},\ }\bibfield  {title} {\bibinfo {title}
  {Kinetic magnetism in triangular moir\'e materials},\ }\href
  {https://doi.org/10.1038/s41586-023-06633-0} {\bibfield  {journal} {\bibinfo
  {journal} {Nature}\ }\textbf {\bibinfo {volume} {623}},\ \bibinfo {pages}
  {509} (\bibinfo {year} {2023})}\BibitemShut {NoStop}%
\bibitem [{\citenamefont {Tang}\ \emph {et~al.}(2020)\citenamefont {Tang},
  \citenamefont {Li}, \citenamefont {Li}, \citenamefont {Xu}, \citenamefont
  {Liu}, \citenamefont {Barmak}, \citenamefont {Watanabe}, \citenamefont
  {Taniguchi}, \citenamefont {MacDonald}, \citenamefont {Shan},\ and\
  \citenamefont {Mak}}]{Tang_2020}%
  \BibitemOpen
  \bibfield  {author} {\bibinfo {author} {\bibfnamefont {Y.}~\bibnamefont
  {Tang}}, \bibinfo {author} {\bibfnamefont {L.}~\bibnamefont {Li}}, \bibinfo
  {author} {\bibfnamefont {T.}~\bibnamefont {Li}}, \bibinfo {author}
  {\bibfnamefont {Y.}~\bibnamefont {Xu}}, \bibinfo {author} {\bibfnamefont
  {S.}~\bibnamefont {Liu}}, \bibinfo {author} {\bibfnamefont {K.}~\bibnamefont
  {Barmak}}, \bibinfo {author} {\bibfnamefont {K.}~\bibnamefont {Watanabe}},
  \bibinfo {author} {\bibfnamefont {T.}~\bibnamefont {Taniguchi}}, \bibinfo
  {author} {\bibfnamefont {A.~H.}\ \bibnamefont {MacDonald}}, \bibinfo {author}
  {\bibfnamefont {J.}~\bibnamefont {Shan}},\ and\ \bibinfo {author}
  {\bibfnamefont {K.~F.}\ \bibnamefont {Mak}},\ }\bibfield  {title} {\bibinfo
  {title} {Simulation of {H}ubbard model physics in {WSe}$_2$/{WS}$_2$ moiré
  superlattices},\ }\href {https://doi.org/10.1038/s41586-020-2085-3}
  {\bibfield  {journal} {\bibinfo  {journal} {Nature}\ }\textbf {\bibinfo
  {volume} {579}},\ \bibinfo {pages} {353} (\bibinfo {year}
  {2020})}\BibitemShut {NoStop}%
\bibitem [{\citenamefont {Xu}\ \emph {et~al.}(2022)\citenamefont {Xu},
  \citenamefont {Kang}, \citenamefont {Watanabe}, \citenamefont {Taniguchi},
  \citenamefont {Mak},\ and\ \citenamefont {Shan}}]{Xu2022}%
  \BibitemOpen
  \bibfield  {author} {\bibinfo {author} {\bibfnamefont {Y.}~\bibnamefont
  {Xu}}, \bibinfo {author} {\bibfnamefont {K.}~\bibnamefont {Kang}}, \bibinfo
  {author} {\bibfnamefont {K.}~\bibnamefont {Watanabe}}, \bibinfo {author}
  {\bibfnamefont {T.}~\bibnamefont {Taniguchi}}, \bibinfo {author}
  {\bibfnamefont {K.~F.}\ \bibnamefont {Mak}},\ and\ \bibinfo {author}
  {\bibfnamefont {J.}~\bibnamefont {Shan}},\ }\bibfield  {title} {\bibinfo
  {title} {{A tunable bilayer Hubbard model in twisted WSe$_2$}},\ }\href
  {https://doi.org/10.1038/s41565-022-01180-7} {\bibfield  {journal} {\bibinfo
  {journal} {Nat. Nanotechnol.}\ }\textbf {\bibinfo {volume} {17}},\ \bibinfo
  {pages} {934} (\bibinfo {year} {2022})}\BibitemShut {NoStop}%
\bibitem [{\citenamefont {Campbell}\ \emph {et~al.}(2022)\citenamefont
  {Campbell}, \citenamefont {Brotons-Gisbert}, \citenamefont {Baek},
  \citenamefont {Vitale}, \citenamefont {Taniguchi}, \citenamefont {Watanabe},
  \citenamefont {Lischner},\ and\ \citenamefont {Gerardot}}]{Campbell_2022}%
  \BibitemOpen
  \bibfield  {author} {\bibinfo {author} {\bibfnamefont {A.~J.}\ \bibnamefont
  {Campbell}}, \bibinfo {author} {\bibfnamefont {M.}~\bibnamefont
  {Brotons-Gisbert}}, \bibinfo {author} {\bibfnamefont {H.}~\bibnamefont
  {Baek}}, \bibinfo {author} {\bibfnamefont {V.}~\bibnamefont {Vitale}},
  \bibinfo {author} {\bibfnamefont {T.}~\bibnamefont {Taniguchi}}, \bibinfo
  {author} {\bibfnamefont {K.}~\bibnamefont {Watanabe}}, \bibinfo {author}
  {\bibfnamefont {J.}~\bibnamefont {Lischner}},\ and\ \bibinfo {author}
  {\bibfnamefont {B.~D.}\ \bibnamefont {Gerardot}},\ }\bibfield  {title}
  {\bibinfo {title} {{Exciton-polarons in the presence of strongly correlated
  electronic states in a MoSe$_2$/WSe$_2$ moir{\ifmmode\acute{e}\else\'{e}\fi}
  superlattice}},\ }\href {https://doi.org/10.1038/s41699-022-00358-w}
  {\bibfield  {journal} {\bibinfo  {journal} {npj 2D Mater. Appl.}\ }\textbf
  {\bibinfo {volume} {6}},\ \bibinfo {pages} {79} (\bibinfo {year}
  {2022})}\BibitemShut {NoStop}%
\bibitem [{\citenamefont {Tang}\ \emph {et~al.}(2023)\citenamefont {Tang},
  \citenamefont {Su}, \citenamefont {Li}, \citenamefont {Xu}, \citenamefont
  {Liu}, \citenamefont {Watanabe}, \citenamefont {Taniguchi}, \citenamefont
  {Hone}, \citenamefont {Jian}, \citenamefont {Xu}, \citenamefont {Mak},\ and\
  \citenamefont {Shan}}]{Tang_2023}%
  \BibitemOpen
  \bibfield  {author} {\bibinfo {author} {\bibfnamefont {Y.}~\bibnamefont
  {Tang}}, \bibinfo {author} {\bibfnamefont {K.}~\bibnamefont {Su}}, \bibinfo
  {author} {\bibfnamefont {L.}~\bibnamefont {Li}}, \bibinfo {author}
  {\bibfnamefont {Y.}~\bibnamefont {Xu}}, \bibinfo {author} {\bibfnamefont
  {S.}~\bibnamefont {Liu}}, \bibinfo {author} {\bibfnamefont {K.}~\bibnamefont
  {Watanabe}}, \bibinfo {author} {\bibfnamefont {T.}~\bibnamefont {Taniguchi}},
  \bibinfo {author} {\bibfnamefont {J.}~\bibnamefont {Hone}}, \bibinfo {author}
  {\bibfnamefont {C.-M.}\ \bibnamefont {Jian}}, \bibinfo {author}
  {\bibfnamefont {C.}~\bibnamefont {Xu}}, \bibinfo {author} {\bibfnamefont
  {K.~F.}\ \bibnamefont {Mak}},\ and\ \bibinfo {author} {\bibfnamefont
  {J.}~\bibnamefont {Shan}},\ }\bibfield  {title} {\bibinfo {title} {{Evidence
  of frustrated magnetic interactions in a Wigner{\textendash}Mott
  insulator}},\ }\href {https://doi.org/10.1038/s41565-022-01309-8} {\bibfield
  {journal} {\bibinfo  {journal} {Nat. Nanotechnol.}\ }\textbf {\bibinfo
  {volume} {18}},\ \bibinfo {pages} {233} (\bibinfo {year} {2023})}\BibitemShut
  {NoStop}%
\bibitem [{\citenamefont {Polovnikov}\ \emph
  {et~al.}(2024{\natexlab{b}})\citenamefont {Polovnikov}, \citenamefont
  {Scherzer}, \citenamefont {Misra}, \citenamefont {Huang}, \citenamefont
  {Mohl}, \citenamefont {Li}, \citenamefont {G\"oser}, \citenamefont
  {F\"orste}, \citenamefont {Bilgin}, \citenamefont {Watanabe}, \citenamefont
  {Taniguchi}, \citenamefont {H\"ogele},\ and\ \citenamefont
  {Baimuratov}}]{Polovnikov_2023}%
  \BibitemOpen
  \bibfield  {author} {\bibinfo {author} {\bibfnamefont {B.}~\bibnamefont
  {Polovnikov}}, \bibinfo {author} {\bibfnamefont {J.}~\bibnamefont
  {Scherzer}}, \bibinfo {author} {\bibfnamefont {S.}~\bibnamefont {Misra}},
  \bibinfo {author} {\bibfnamefont {X.}~\bibnamefont {Huang}}, \bibinfo
  {author} {\bibfnamefont {C.}~\bibnamefont {Mohl}}, \bibinfo {author}
  {\bibfnamefont {Z.}~\bibnamefont {Li}}, \bibinfo {author} {\bibfnamefont
  {J.}~\bibnamefont {G\"oser}}, \bibinfo {author} {\bibfnamefont
  {J.}~\bibnamefont {F\"orste}}, \bibinfo {author} {\bibfnamefont
  {I.}~\bibnamefont {Bilgin}}, \bibinfo {author} {\bibfnamefont
  {K.}~\bibnamefont {Watanabe}}, \bibinfo {author} {\bibfnamefont
  {T.}~\bibnamefont {Taniguchi}}, \bibinfo {author} {\bibfnamefont
  {A.}~\bibnamefont {H\"ogele}},\ and\ \bibinfo {author} {\bibfnamefont
  {A.~S.}\ \bibnamefont {Baimuratov}},\ }\bibfield  {title} {\bibinfo {title}
  {{Field-Induced Hybridization of Moir\'e Excitons in MoSe$_{2}$/WS$_{2}$
  Heterobilayers}},\ }\href {https://doi.org/10.1103/PhysRevLett.132.076902}
  {\bibfield  {journal} {\bibinfo  {journal} {Phys. Rev. Lett.}\ }\textbf
  {\bibinfo {volume} {132}},\ \bibinfo {pages} {076902} (\bibinfo {year}
  {2024}{\natexlab{b}})}\BibitemShut {NoStop}%
\bibitem [{\citenamefont {Schneider}\ \emph {et~al.}(2018)\citenamefont
  {Schneider}, \citenamefont {Glazov}, \citenamefont {Korn}, \citenamefont
  {Höfling},\ and\ \citenamefont {Urbaszek}}]{Schneider_2018}%
  \BibitemOpen
  \bibfield  {author} {\bibinfo {author} {\bibfnamefont {C.}~\bibnamefont
  {Schneider}}, \bibinfo {author} {\bibfnamefont {M.}~\bibnamefont {Glazov}},
  \bibinfo {author} {\bibfnamefont {T.}~\bibnamefont {Korn}}, \bibinfo {author}
  {\bibfnamefont {S.}~\bibnamefont {Höfling}},\ and\ \bibinfo {author}
  {\bibfnamefont {B.}~\bibnamefont {Urbaszek}},\ }\bibfield  {title} {\bibinfo
  {title} {Two-dimensional semiconductors in the regime of strong light-matter
  coupling},\ }\href {http://dx.doi.org/10.1038/s41467-018-04866-6} {\bibfield
  {journal} {\bibinfo  {journal} {Nat. Commun.}\ }\textbf {\bibinfo {volume}
  {9}},\ \bibinfo {pages} {2695} (\bibinfo {year} {2018})}\BibitemShut
  {NoStop}%
\bibitem [{\citenamefont {Drawer}\ \emph {et~al.}(2023)\citenamefont {Drawer},
  \citenamefont {Mitryakhin}, \citenamefont {Shan}, \citenamefont {Stephan},
  \citenamefont {Gittinger}, \citenamefont {Lackner}, \citenamefont {Han},
  \citenamefont {Leibeling}, \citenamefont {Eilenberger}, \citenamefont
  {Banerjee}, \citenamefont {Tongay}, \citenamefont {Watanabe}, \citenamefont
  {Taniguchi}, \citenamefont {Lienau}, \citenamefont {Silies}, \citenamefont
  {Anton-Solanas}, \citenamefont {Esmann},\ and\ \citenamefont
  {Schneider}}]{Drawer_2023}%
  \BibitemOpen
  \bibfield  {author} {\bibinfo {author} {\bibfnamefont {J.-C.}\ \bibnamefont
  {Drawer}}, \bibinfo {author} {\bibfnamefont {V.~N.}\ \bibnamefont
  {Mitryakhin}}, \bibinfo {author} {\bibfnamefont {H.}~\bibnamefont {Shan}},
  \bibinfo {author} {\bibfnamefont {S.}~\bibnamefont {Stephan}}, \bibinfo
  {author} {\bibfnamefont {M.}~\bibnamefont {Gittinger}}, \bibinfo {author}
  {\bibfnamefont {L.}~\bibnamefont {Lackner}}, \bibinfo {author} {\bibfnamefont
  {B.}~\bibnamefont {Han}}, \bibinfo {author} {\bibfnamefont {G.}~\bibnamefont
  {Leibeling}}, \bibinfo {author} {\bibfnamefont {F.}~\bibnamefont
  {Eilenberger}}, \bibinfo {author} {\bibfnamefont {R.}~\bibnamefont
  {Banerjee}}, \bibinfo {author} {\bibfnamefont {S.}~\bibnamefont {Tongay}},
  \bibinfo {author} {\bibfnamefont {K.}~\bibnamefont {Watanabe}}, \bibinfo
  {author} {\bibfnamefont {T.}~\bibnamefont {Taniguchi}}, \bibinfo {author}
  {\bibfnamefont {C.}~\bibnamefont {Lienau}}, \bibinfo {author} {\bibfnamefont
  {M.}~\bibnamefont {Silies}}, \bibinfo {author} {\bibfnamefont
  {C.}~\bibnamefont {Anton-Solanas}}, \bibinfo {author} {\bibfnamefont
  {M.}~\bibnamefont {Esmann}},\ and\ \bibinfo {author} {\bibfnamefont
  {C.}~\bibnamefont {Schneider}},\ }\bibfield  {title} {\bibinfo {title}
  {Monolayer-based single-photon source in a liquid-helium-free open cavity
  featuring 65 \% brightness and quantum coherence},\ }\href
  {http://dx.doi.org/10.1021/acs.nanolett.3c02584} {\bibfield  {journal}
  {\bibinfo  {journal} {Nano Lett.}\ }\textbf {\bibinfo {volume} {23}},\
  \bibinfo {pages} {8683–8689} (\bibinfo {year} {2023})}\BibitemShut
  {NoStop}%
\bibitem [{\citenamefont {Weisbuch}\ \emph {et~al.}(1993)\citenamefont
  {Weisbuch}, \citenamefont {Nishioka}, \citenamefont {Ishikawa},\ and\
  \citenamefont {Arakawa}}]{Weisbuch_1993}%
  \BibitemOpen
  \bibfield  {author} {\bibinfo {author} {\bibfnamefont {C.}~\bibnamefont
  {Weisbuch}}, \bibinfo {author} {\bibfnamefont {M.}~\bibnamefont {Nishioka}},
  \bibinfo {author} {\bibfnamefont {A.}~\bibnamefont {Ishikawa}},\ and\
  \bibinfo {author} {\bibfnamefont {Y.}~\bibnamefont {Arakawa}},\ }\bibfield
  {title} {\bibinfo {title} {Observation of the coupled exciton-photon mode
  splitting in a semiconductor quantum microcavity},\ }\href
  {https://doi.org/10.1103/PhysRevLett.69.3314} {\bibfield  {journal} {\bibinfo
   {journal} {Phys. Rev. Lett.}\ }\textbf {\bibinfo {volume} {69}},\ \bibinfo
  {pages} {3314} (\bibinfo {year} {1993})}\BibitemShut {NoStop}%
\bibitem [{\citenamefont {Kavokin}\ \emph {et~al.}(2011)\citenamefont
  {Kavokin}, \citenamefont {Baumberg}, \citenamefont {Malpuech},\ and\
  \citenamefont {Laussy}}]{Kavokin_2011}%
  \BibitemOpen
  \bibfield  {author} {\bibinfo {author} {\bibfnamefont {A.}~\bibnamefont
  {Kavokin}}, \bibinfo {author} {\bibfnamefont {J.}~\bibnamefont {Baumberg}},
  \bibinfo {author} {\bibfnamefont {G.}~\bibnamefont {Malpuech}},\ and\
  \bibinfo {author} {\bibfnamefont {F.}~\bibnamefont {Laussy}},\ }\href
  {https://books.google.de/books?id=2g7wHcMcaJ0C} {\emph {\bibinfo {title}
  {Microcavities}}}\ (\bibinfo  {publisher} {Oxford university press},\
  \bibinfo {year} {2011})\BibitemShut {NoStop}%
\bibitem [{\citenamefont {Muñoz-Matutano}\ \emph {et~al.}(2019)\citenamefont
  {Muñoz-Matutano}, \citenamefont {Wood}, \citenamefont {Johnsson},
  \citenamefont {Vidal}, \citenamefont {Baragiola}, \citenamefont {Reinhard},
  \citenamefont {Lemaître}, \citenamefont {Bloch}, \citenamefont {Amo},
  \citenamefont {Nogues}, \citenamefont {Besga}, \citenamefont {Richard},\ and\
  \citenamefont {Volz}}]{MunozMatutano_2019}%
  \BibitemOpen
  \bibfield  {author} {\bibinfo {author} {\bibfnamefont {G.}~\bibnamefont
  {Muñoz-Matutano}}, \bibinfo {author} {\bibfnamefont {A.}~\bibnamefont
  {Wood}}, \bibinfo {author} {\bibfnamefont {M.}~\bibnamefont {Johnsson}},
  \bibinfo {author} {\bibfnamefont {X.}~\bibnamefont {Vidal}}, \bibinfo
  {author} {\bibfnamefont {B.~Q.}\ \bibnamefont {Baragiola}}, \bibinfo {author}
  {\bibfnamefont {A.}~\bibnamefont {Reinhard}}, \bibinfo {author}
  {\bibfnamefont {A.}~\bibnamefont {Lemaître}}, \bibinfo {author}
  {\bibfnamefont {J.}~\bibnamefont {Bloch}}, \bibinfo {author} {\bibfnamefont
  {A.}~\bibnamefont {Amo}}, \bibinfo {author} {\bibfnamefont {G.}~\bibnamefont
  {Nogues}}, \bibinfo {author} {\bibfnamefont {B.}~\bibnamefont {Besga}},
  \bibinfo {author} {\bibfnamefont {M.}~\bibnamefont {Richard}},\ and\ \bibinfo
  {author} {\bibfnamefont {T.}~\bibnamefont {Volz}},\ }\bibfield  {title}
  {\bibinfo {title} {Emergence of quantum correlations from interacting
  fibre-cavity polaritons},\ }\href {https://doi.org/10.1038/s41563-019-0281-z}
  {\bibfield  {journal} {\bibinfo  {journal} {Nat. Mater.}\ }\textbf {\bibinfo
  {volume} {18}},\ \bibinfo {pages} {213–218} (\bibinfo {year}
  {2019})}\BibitemShut {NoStop}%
\bibitem [{\citenamefont {Delteil}\ \emph {et~al.}(2019)\citenamefont
  {Delteil}, \citenamefont {Fink}, \citenamefont {Schade}, \citenamefont
  {Höfling}, \citenamefont {Schneider},\ and\ \citenamefont
  {İmamoğlu}}]{Delteil_2019}%
  \BibitemOpen
  \bibfield  {author} {\bibinfo {author} {\bibfnamefont {A.}~\bibnamefont
  {Delteil}}, \bibinfo {author} {\bibfnamefont {T.}~\bibnamefont {Fink}},
  \bibinfo {author} {\bibfnamefont {A.}~\bibnamefont {Schade}}, \bibinfo
  {author} {\bibfnamefont {S.}~\bibnamefont {Höfling}}, \bibinfo {author}
  {\bibfnamefont {C.}~\bibnamefont {Schneider}},\ and\ \bibinfo {author}
  {\bibfnamefont {A.}~\bibnamefont {İmamoğlu}},\ }\bibfield  {title}
  {\bibinfo {title} {Towards polariton blockade of confined
  exciton–polaritons},\ }\href {https://doi.org/10.1038/s41563-019-0282-y}
  {\bibfield  {journal} {\bibinfo  {journal} {Nat. Mater.}\ }\textbf {\bibinfo
  {volume} {18}},\ \bibinfo {pages} {219–222} (\bibinfo {year}
  {2019})}\BibitemShut {NoStop}%
\bibitem [{\citenamefont {Kasprzak}\ \emph {et~al.}(2006)\citenamefont
  {Kasprzak}, \citenamefont {Richard}, \citenamefont {Kundermann},
  \citenamefont {Baas}, \citenamefont {Jeambrun}, \citenamefont {Keeling},
  \citenamefont {Marchetti}, \citenamefont {Szymańska}, \citenamefont
  {André}, \citenamefont {Staehli}, \citenamefont {Savona}, \citenamefont
  {Littlewood}, \citenamefont {Deveaud},\ and\ \citenamefont
  {Le~Si}}]{Kasprzak_2006}%
  \BibitemOpen
  \bibfield  {author} {\bibinfo {author} {\bibfnamefont {J.}~\bibnamefont
  {Kasprzak}}, \bibinfo {author} {\bibfnamefont {M.}~\bibnamefont {Richard}},
  \bibinfo {author} {\bibfnamefont {S.}~\bibnamefont {Kundermann}}, \bibinfo
  {author} {\bibfnamefont {A.}~\bibnamefont {Baas}}, \bibinfo {author}
  {\bibfnamefont {P.}~\bibnamefont {Jeambrun}}, \bibinfo {author}
  {\bibfnamefont {J.}~\bibnamefont {Keeling}}, \bibinfo {author} {\bibfnamefont
  {F.~M.}\ \bibnamefont {Marchetti}}, \bibinfo {author} {\bibfnamefont
  {M.}~\bibnamefont {Szymańska}}, \bibinfo {author} {\bibfnamefont
  {R.}~\bibnamefont {André}}, \bibinfo {author} {\bibfnamefont
  {J.}~\bibnamefont {Staehli}}, \bibinfo {author} {\bibfnamefont
  {V.}~\bibnamefont {Savona}}, \bibinfo {author} {\bibfnamefont
  {P.}~\bibnamefont {Littlewood}}, \bibinfo {author} {\bibfnamefont
  {B.}~\bibnamefont {Deveaud}},\ and\ \bibinfo {author} {\bibfnamefont
  {D.}~\bibnamefont {Le~Si}},\ }\bibfield  {title} {\bibinfo {title}
  {{Bose{\textendash}Einstein condensation of exciton polaritons}},\ }\href
  {https://doi.org/10.1038/nature05131} {\bibfield  {journal} {\bibinfo
  {journal} {Nature}\ }\textbf {\bibinfo {volume} {443}},\ \bibinfo {pages}
  {409} (\bibinfo {year} {2006})}\BibitemShut {NoStop}%
\bibitem [{\citenamefont {Anton-Solanas}\ \emph {et~al.}(2021)\citenamefont
  {Anton-Solanas}, \citenamefont {Waldherr}, \citenamefont {Klaas},
  \citenamefont {Suchomel}, \citenamefont {Harder}, \citenamefont {Cai},
  \citenamefont {Sedov}, \citenamefont {Klembt}, \citenamefont {Kavokin},
  \citenamefont {Tongay}, \citenamefont {Watanabe}, \citenamefont {Taniguchi},
  \citenamefont {Höfling},\ and\ \citenamefont
  {Schneider}}]{Anton-Solanas_2021}%
  \BibitemOpen
  \bibfield  {author} {\bibinfo {author} {\bibfnamefont {C.}~\bibnamefont
  {Anton-Solanas}}, \bibinfo {author} {\bibfnamefont {M.}~\bibnamefont
  {Waldherr}}, \bibinfo {author} {\bibfnamefont {M.}~\bibnamefont {Klaas}},
  \bibinfo {author} {\bibfnamefont {H.}~\bibnamefont {Suchomel}}, \bibinfo
  {author} {\bibfnamefont {T.}~\bibnamefont {Harder}}, \bibinfo {author}
  {\bibfnamefont {H.}~\bibnamefont {Cai}}, \bibinfo {author} {\bibfnamefont
  {E.}~\bibnamefont {Sedov}}, \bibinfo {author} {\bibfnamefont
  {S.}~\bibnamefont {Klembt}}, \bibinfo {author} {\bibfnamefont
  {A.}~\bibnamefont {Kavokin}}, \bibinfo {author} {\bibfnamefont
  {S.}~\bibnamefont {Tongay}}, \bibinfo {author} {\bibfnamefont
  {K.}~\bibnamefont {Watanabe}}, \bibinfo {author} {\bibfnamefont
  {T.}~\bibnamefont {Taniguchi}}, \bibinfo {author} {\bibfnamefont
  {S.}~\bibnamefont {Höfling}},\ and\ \bibinfo {author} {\bibfnamefont
  {C.}~\bibnamefont {Schneider}},\ }\bibfield  {title} {\bibinfo {title}
  {Bosonic condensation of exciton–polaritons in an atomically thin
  crystal},\ }\href {https://doi.org/10.1038/s41563-021-01000-8} {\bibfield
  {journal} {\bibinfo  {journal} {Nat. Mater.}\ }\textbf {\bibinfo {volume}
  {20}},\ \bibinfo {pages} {1233} (\bibinfo {year} {2021})}\BibitemShut
  {NoStop}%
\bibitem [{\citenamefont {K\'ena-Cohen}\ and\ \citenamefont
  {Forrest}(2010)}]{Kena-Cohen_2010}%
  \BibitemOpen
  \bibfield  {author} {\bibinfo {author} {\bibfnamefont {S.}~\bibnamefont
  {K\'ena-Cohen}}\ and\ \bibinfo {author} {\bibfnamefont {S.}~\bibnamefont
  {Forrest}},\ }\bibfield  {title} {\bibinfo {title} {Room-temperature
  polariton lasing in an organic single-crystal microcavity},\ }\href
  {https://doi.org/10.1038/nphoton.2010.86} {\bibfield  {journal} {\bibinfo
  {journal} {Nat. Photon.}\ }\textbf {\bibinfo {volume} {4}},\ \bibinfo {pages}
  {371} (\bibinfo {year} {2010})}\BibitemShut {NoStop}%
\bibitem [{\citenamefont {Amo}\ \emph {et~al.}(2011)\citenamefont {Amo},
  \citenamefont {Pigeon}, \citenamefont {Sanvitto}, \citenamefont {Sala},
  \citenamefont {Hivet}, \citenamefont {Carusotto}, \citenamefont {Pisanello},
  \citenamefont {Lem{\'{e}}nager}, \citenamefont {Houdr{\'{e}}}, \citenamefont
  {Giacobino}, \citenamefont {Ciuti},\ and\ \citenamefont
  {Bramati}}]{Amo_2011}%
  \BibitemOpen
  \bibfield  {author} {\bibinfo {author} {\bibfnamefont {A.}~\bibnamefont
  {Amo}}, \bibinfo {author} {\bibfnamefont {S.}~\bibnamefont {Pigeon}},
  \bibinfo {author} {\bibfnamefont {D.}~\bibnamefont {Sanvitto}}, \bibinfo
  {author} {\bibfnamefont {V.~G.}\ \bibnamefont {Sala}}, \bibinfo {author}
  {\bibfnamefont {R.}~\bibnamefont {Hivet}}, \bibinfo {author} {\bibfnamefont
  {I.}~\bibnamefont {Carusotto}}, \bibinfo {author} {\bibfnamefont
  {F.}~\bibnamefont {Pisanello}}, \bibinfo {author} {\bibfnamefont
  {G.}~\bibnamefont {Lem{\'{e}}nager}}, \bibinfo {author} {\bibfnamefont
  {R.}~\bibnamefont {Houdr{\'{e}}}}, \bibinfo {author} {\bibfnamefont
  {E.}~\bibnamefont {Giacobino}}, \bibinfo {author} {\bibfnamefont
  {C.}~\bibnamefont {Ciuti}},\ and\ \bibinfo {author} {\bibfnamefont
  {A.}~\bibnamefont {Bramati}},\ }\bibfield  {title} {\bibinfo {title}
  {Polariton superfluids reveal quantum hydrodynamic solitons},\ }\href
  {https://doi.org/10.1126/science.1202307} {\bibfield  {journal} {\bibinfo
  {journal} {Science}\ }\textbf {\bibinfo {volume} {332}},\ \bibinfo {pages}
  {1167} (\bibinfo {year} {2011})}\BibitemShut {NoStop}%
\bibitem [{\citenamefont {Gu}\ \emph {et~al.}(2021)\citenamefont {Gu},
  \citenamefont {Walther}, \citenamefont {Waldecker}, \citenamefont {Rhodes},
  \citenamefont {Raja}, \citenamefont {Hone}, \citenamefont {Heinz},
  \citenamefont {Kéna-Cohen}, \citenamefont {Pohl},\ and\ \citenamefont
  {Menon}}]{Gu2021}%
  \BibitemOpen
  \bibfield  {author} {\bibinfo {author} {\bibfnamefont {J.}~\bibnamefont
  {Gu}}, \bibinfo {author} {\bibfnamefont {V.}~\bibnamefont {Walther}},
  \bibinfo {author} {\bibfnamefont {L.}~\bibnamefont {Waldecker}}, \bibinfo
  {author} {\bibfnamefont {D.}~\bibnamefont {Rhodes}}, \bibinfo {author}
  {\bibfnamefont {A.}~\bibnamefont {Raja}}, \bibinfo {author} {\bibfnamefont
  {J.~C.}\ \bibnamefont {Hone}}, \bibinfo {author} {\bibfnamefont {T.~F.}\
  \bibnamefont {Heinz}}, \bibinfo {author} {\bibfnamefont {S.}~\bibnamefont
  {Kéna-Cohen}}, \bibinfo {author} {\bibfnamefont {T.}~\bibnamefont {Pohl}},\
  and\ \bibinfo {author} {\bibfnamefont {V.~M.}\ \bibnamefont {Menon}},\
  }\bibfield  {title} {\bibinfo {title} {{Enhanced nonlinear interaction of
  polaritons via excitonic Rydberg states in monolayer WSe$_2$}},\ }\href
  {https://doi.org/10.1038/s41467-021-22537-x} {\bibfield  {journal} {\bibinfo
  {journal} {Nat. Commun.}\ }\textbf {\bibinfo {volume} {12}},\ \bibinfo
  {pages} {2269} (\bibinfo {year} {2021})}\BibitemShut {NoStop}%
\bibitem [{\citenamefont {Sidler}\ \emph {et~al.}(2016)\citenamefont {Sidler},
  \citenamefont {Back}, \citenamefont {Cotlet}, \citenamefont {Srivastava},
  \citenamefont {Fink}, \citenamefont {Kroner}, \citenamefont {Demler},\ and\
  \citenamefont {Imamoglu}}]{Sidler_2017}%
  \BibitemOpen
  \bibfield  {author} {\bibinfo {author} {\bibfnamefont {M.}~\bibnamefont
  {Sidler}}, \bibinfo {author} {\bibfnamefont {P.}~\bibnamefont {Back}},
  \bibinfo {author} {\bibfnamefont {O.}~\bibnamefont {Cotlet}}, \bibinfo
  {author} {\bibfnamefont {A.}~\bibnamefont {Srivastava}}, \bibinfo {author}
  {\bibfnamefont {T.}~\bibnamefont {Fink}}, \bibinfo {author} {\bibfnamefont
  {M.}~\bibnamefont {Kroner}}, \bibinfo {author} {\bibfnamefont
  {E.}~\bibnamefont {Demler}},\ and\ \bibinfo {author} {\bibfnamefont
  {A.}~\bibnamefont {Imamoglu}},\ }\bibfield  {title} {\bibinfo {title} {Fermi
  polaron-polaritons in charge-tunable atomically thin semiconductors},\ }\href
  {https://doi.org/10.1038/nphys3949} {\bibfield  {journal} {\bibinfo
  {journal} {Nat. Phys.}\ }\textbf {\bibinfo {volume} {13}},\ \bibinfo {pages}
  {255} (\bibinfo {year} {2016})}\BibitemShut {NoStop}%
\bibitem [{\citenamefont {Efimkin}\ and\ \citenamefont
  {Macdonald}(2017)}]{Efimkin_2017}%
  \BibitemOpen
  \bibfield  {author} {\bibinfo {author} {\bibfnamefont {D.}~\bibnamefont
  {Efimkin}}\ and\ \bibinfo {author} {\bibfnamefont {A.}~\bibnamefont
  {Macdonald}},\ }\bibfield  {title} {\bibinfo {title} {Many-body theory of
  trion absorption features in two-dimensional semiconductors},\ }\href
  {https://doi.org/10.1103/PhysRevB.95.035417} {\bibfield  {journal} {\bibinfo
  {journal} {Phys. Rev. B}\ }\textbf {\bibinfo {volume} {95}},\ \bibinfo
  {pages} {035417} (\bibinfo {year} {2017})}\BibitemShut {NoStop}%
\bibitem [{\citenamefont {Back}\ \emph {et~al.}(2017)\citenamefont {Back},
  \citenamefont {Sidler}, \citenamefont {Cotlet}, \citenamefont {Srivastava},
  \citenamefont {Takemura}, \citenamefont {Kroner},\ and\ \citenamefont
  {Imamo{\ifmmode\breve{g}\else\u{g}\fi}lu}}]{Back_2017}%
  \BibitemOpen
  \bibfield  {author} {\bibinfo {author} {\bibfnamefont {P.}~\bibnamefont
  {Back}}, \bibinfo {author} {\bibfnamefont {M.}~\bibnamefont {Sidler}},
  \bibinfo {author} {\bibfnamefont {O.}~\bibnamefont {Cotlet}}, \bibinfo
  {author} {\bibfnamefont {A.}~\bibnamefont {Srivastava}}, \bibinfo {author}
  {\bibfnamefont {N.}~\bibnamefont {Takemura}}, \bibinfo {author}
  {\bibfnamefont {M.}~\bibnamefont {Kroner}},\ and\ \bibinfo {author}
  {\bibfnamefont {A.}~\bibnamefont {Imamo{\ifmmode\breve{g}\else\u{g}\fi}lu}},\
  }\bibfield  {title} {\bibinfo {title} {{Giant Paramagnetism-Induced Valley
  Polarization of Electrons in Charge-Tunable Monolayer
  ${\mathrm{MoSe}}_{2}$}},\ }\href
  {https://doi.org/10.1103/PhysRevLett.118.237404} {\bibfield  {journal}
  {\bibinfo  {journal} {Phys. Rev. Lett.}\ }\textbf {\bibinfo {volume} {118}},\
  \bibinfo {pages} {237404} (\bibinfo {year} {2017})}\BibitemShut {NoStop}%
\bibitem [{\citenamefont {Lyons}\ \emph {et~al.}(2022)\citenamefont {Lyons},
  \citenamefont {Gillard}, \citenamefont {Leblanc}, \citenamefont {Puebla},
  \citenamefont {Solnyshkov}, \citenamefont {Klompmaker}, \citenamefont
  {Akimov}, \citenamefont {Louca}, \citenamefont {Muduli}, \citenamefont
  {Genco}, \citenamefont {Bayer}, \citenamefont {Otani}, \citenamefont
  {Malpuech},\ and\ \citenamefont {Tartakovskii}}]{Lyons_2022}%
  \BibitemOpen
  \bibfield  {author} {\bibinfo {author} {\bibfnamefont {T.~P.}\ \bibnamefont
  {Lyons}}, \bibinfo {author} {\bibfnamefont {D.~J.}\ \bibnamefont {Gillard}},
  \bibinfo {author} {\bibfnamefont {C.}~\bibnamefont {Leblanc}}, \bibinfo
  {author} {\bibfnamefont {J.}~\bibnamefont {Puebla}}, \bibinfo {author}
  {\bibfnamefont {D.~D.}\ \bibnamefont {Solnyshkov}}, \bibinfo {author}
  {\bibfnamefont {L.}~\bibnamefont {Klompmaker}}, \bibinfo {author}
  {\bibfnamefont {I.~A.}\ \bibnamefont {Akimov}}, \bibinfo {author}
  {\bibfnamefont {C.}~\bibnamefont {Louca}}, \bibinfo {author} {\bibfnamefont
  {P.}~\bibnamefont {Muduli}}, \bibinfo {author} {\bibfnamefont
  {A.}~\bibnamefont {Genco}}, \bibinfo {author} {\bibfnamefont
  {M.}~\bibnamefont {Bayer}}, \bibinfo {author} {\bibfnamefont
  {Y.}~\bibnamefont {Otani}}, \bibinfo {author} {\bibfnamefont
  {G.}~\bibnamefont {Malpuech}},\ and\ \bibinfo {author} {\bibfnamefont
  {A.~I.}\ \bibnamefont {Tartakovskii}},\ }\bibfield  {title} {\bibinfo {title}
  {{Giant effective Zeeman splitting in a monolayer semiconductor realized by
  spin-selective strong light{\textendash}matter coupling}},\ }\href
  {https://doi.org/10.1038/s41566-022-01025-8} {\bibfield  {journal} {\bibinfo
  {journal} {Nat. Photon.}\ }\textbf {\bibinfo {volume} {16}},\ \bibinfo
  {pages} {632} (\bibinfo {year} {2022})}\BibitemShut {NoStop}%
\bibitem [{\citenamefont {Morera}\ \emph {et~al.}(2023)\citenamefont {Morera},
  \citenamefont {Kan{\ifmmode\acute{a}\else\'{a}\fi}sz-Nagy}, \citenamefont
  {Smolenski}, \citenamefont {Ciorciaro}, \citenamefont
  {Imamo{\ifmmode\breve{g}\else\u{g}\fi}lu},\ and\ \citenamefont
  {Demler}}]{Morera_2023}%
  \BibitemOpen
  \bibfield  {author} {\bibinfo {author} {\bibfnamefont {I.}~\bibnamefont
  {Morera}}, \bibinfo {author} {\bibfnamefont {M.}~\bibnamefont
  {Kan{\ifmmode\acute{a}\else\'{a}\fi}sz-Nagy}}, \bibinfo {author}
  {\bibfnamefont {T.}~\bibnamefont {Smolenski}}, \bibinfo {author}
  {\bibfnamefont {L.}~\bibnamefont {Ciorciaro}}, \bibinfo {author}
  {\bibfnamefont {A.}~\bibnamefont {Imamo{\ifmmode\breve{g}\else\u{g}\fi}lu}},\
  and\ \bibinfo {author} {\bibfnamefont {E.}~\bibnamefont {Demler}},\
  }\bibfield  {title} {\bibinfo {title} {{High-temperature kinetic magnetism in
  triangular lattices}},\ }\href
  {https://doi.org/10.1103/PhysRevResearch.5.L022048} {\bibfield  {journal}
  {\bibinfo  {journal} {Phys. Rev. Res.}\ }\textbf {\bibinfo {volume} {5}},\
  \bibinfo {pages} {L022048} (\bibinfo {year} {2023})}\BibitemShut {NoStop}%
\bibitem [{\citenamefont {Zhang}\ \emph {et~al.}(2021)\citenamefont {Zhang},
  \citenamefont {Wu}, \citenamefont {Hou}, \citenamefont {Zhang}, \citenamefont
  {Chou}, \citenamefont {Watanabe}, \citenamefont {Taniguchi}, \citenamefont
  {Forrest},\ and\ \citenamefont {Deng}}]{Zhang_2021}%
  \BibitemOpen
  \bibfield  {author} {\bibinfo {author} {\bibfnamefont {L.}~\bibnamefont
  {Zhang}}, \bibinfo {author} {\bibfnamefont {F.}~\bibnamefont {Wu}}, \bibinfo
  {author} {\bibfnamefont {S.}~\bibnamefont {Hou}}, \bibinfo {author}
  {\bibfnamefont {Z.}~\bibnamefont {Zhang}}, \bibinfo {author} {\bibfnamefont
  {Y.-H.}\ \bibnamefont {Chou}}, \bibinfo {author} {\bibfnamefont
  {K.}~\bibnamefont {Watanabe}}, \bibinfo {author} {\bibfnamefont
  {T.}~\bibnamefont {Taniguchi}}, \bibinfo {author} {\bibfnamefont {S.~R.}\
  \bibnamefont {Forrest}},\ and\ \bibinfo {author} {\bibfnamefont
  {H.}~\bibnamefont {Deng}},\ }\bibfield  {title} {\bibinfo {title} {{Van der
  Waals heterostructure polaritons with moir{\'{e}}-induced nonlinearity}},\
  }\href {https://doi.org/10.1038/s41586-021-03228-5} {\bibfield  {journal}
  {\bibinfo  {journal} {Nature}\ }\textbf {\bibinfo {volume} {591}},\ \bibinfo
  {pages} {61} (\bibinfo {year} {2021})}\BibitemShut {NoStop}%
\bibitem [{\citenamefont {Bilgin}\ \emph {et~al.}(2015)\citenamefont {Bilgin},
  \citenamefont {Liu}, \citenamefont {Vargas}, \citenamefont {Winchester},
  \citenamefont {Man}, \citenamefont {Upmanyu}, \citenamefont {Dani},
  \citenamefont {Gupta}, \citenamefont {Talapatra}, \citenamefont {Mohite},\
  and\ \citenamefont {Kar}}]{Bilgin_2015}%
  \BibitemOpen
  \bibfield  {author} {\bibinfo {author} {\bibfnamefont {I.}~\bibnamefont
  {Bilgin}}, \bibinfo {author} {\bibfnamefont {F.}~\bibnamefont {Liu}},
  \bibinfo {author} {\bibfnamefont {A.}~\bibnamefont {Vargas}}, \bibinfo
  {author} {\bibfnamefont {A.}~\bibnamefont {Winchester}}, \bibinfo {author}
  {\bibfnamefont {M.~K.~L.}\ \bibnamefont {Man}}, \bibinfo {author}
  {\bibfnamefont {M.}~\bibnamefont {Upmanyu}}, \bibinfo {author} {\bibfnamefont
  {K.~M.}\ \bibnamefont {Dani}}, \bibinfo {author} {\bibfnamefont
  {G.}~\bibnamefont {Gupta}}, \bibinfo {author} {\bibfnamefont
  {S.}~\bibnamefont {Talapatra}}, \bibinfo {author} {\bibfnamefont {A.~D.}\
  \bibnamefont {Mohite}},\ and\ \bibinfo {author} {\bibfnamefont
  {S.}~\bibnamefont {Kar}},\ }\bibfield  {title} {\bibinfo {title} {Chemical
  vapor deposition synthesized atomically thin molybdenum disulfide with
  optoelectronic-grade crystalline quality},\ }\href
  {https://doi.org/10.1021/acsnano.5b02019} {\bibfield  {journal} {\bibinfo
  {journal} {ACS Nano}\ }\textbf {\bibinfo {volume} {9}},\ \bibinfo {pages}
  {8822} (\bibinfo {year} {2015})}\BibitemShut {NoStop}%
\bibitem [{\citenamefont {Wu}\ \emph {et~al.}(2017)\citenamefont {Wu},
  \citenamefont {Lovorn},\ and\ \citenamefont {MacDonald}}]{Wu_2017}%
  \BibitemOpen
  \bibfield  {author} {\bibinfo {author} {\bibfnamefont {F.}~\bibnamefont
  {Wu}}, \bibinfo {author} {\bibfnamefont {T.}~\bibnamefont {Lovorn}},\ and\
  \bibinfo {author} {\bibfnamefont {A.~H.}\ \bibnamefont {MacDonald}},\
  }\bibfield  {title} {\bibinfo {title} {{Topological Exciton Bands in Moir\'e
  Heterojunctions}},\ }\href {https://doi.org/10.1103/PhysRevLett.118.147401}
  {\bibfield  {journal} {\bibinfo  {journal} {Phys. Rev. Lett.}\ }\textbf
  {\bibinfo {volume} {118}},\ \bibinfo {pages} {147401} (\bibinfo {year}
  {2017})}\BibitemShut {NoStop}%
\bibitem [{\citenamefont {Wu}\ \emph {et~al.}(2018{\natexlab{b}})\citenamefont
  {Wu}, \citenamefont {Lovorn},\ and\ \citenamefont {MacDonald}}]{Wu_2018_jan}%
  \BibitemOpen
  \bibfield  {author} {\bibinfo {author} {\bibfnamefont {F.}~\bibnamefont
  {Wu}}, \bibinfo {author} {\bibfnamefont {T.}~\bibnamefont {Lovorn}},\ and\
  \bibinfo {author} {\bibfnamefont {A.~H.}\ \bibnamefont {MacDonald}},\
  }\bibfield  {title} {\bibinfo {title} {Theory of optical absorption by
  interlayer excitons in transition metal dichalcogenide heterobilayers},\
  }\href {https://doi.org/10.1103/PhysRevB.97.035306} {\bibfield  {journal}
  {\bibinfo  {journal} {Phys. Rev. B}\ }\textbf {\bibinfo {volume} {97}},\
  \bibinfo {pages} {035306} (\bibinfo {year} {2018}{\natexlab{b}})}\BibitemShut
  {NoStop}%
\bibitem [{\citenamefont {Tong}\ \emph {et~al.}(2020)\citenamefont {Tong},
  \citenamefont {Chen}, \citenamefont {Xiao}, \citenamefont {Yu},\ and\
  \citenamefont {Yao}}]{Tong_2020}%
  \BibitemOpen
  \bibfield  {author} {\bibinfo {author} {\bibfnamefont {Q.}~\bibnamefont
  {Tong}}, \bibinfo {author} {\bibfnamefont {M.}~\bibnamefont {Chen}}, \bibinfo
  {author} {\bibfnamefont {F.}~\bibnamefont {Xiao}}, \bibinfo {author}
  {\bibfnamefont {H.}~\bibnamefont {Yu}},\ and\ \bibinfo {author}
  {\bibfnamefont {W.}~\bibnamefont {Yao}},\ }\bibfield  {title} {\bibinfo
  {title} {{Interferences of electrostatic moir{\ifmmode\acute{e}\else\'{e}\fi}
  potentials and bichromatic superlattices of electrons and excitons in
  transition metal dichalcogenides}},\ }\href
  {https://doi.org/10.1088/2053-1583/abd006} {\bibfield  {journal} {\bibinfo
  {journal} {2D Mater.}\ }\textbf {\bibinfo {volume} {8}},\ \bibinfo {pages}
  {025007} (\bibinfo {year} {2020})}\BibitemShut {NoStop}%
\bibitem [{\citenamefont {Alexeev}\ \emph {et~al.}(2019)\citenamefont
  {Alexeev}, \citenamefont {Ruiz-Tijerina}, \citenamefont {Danovich},
  \citenamefont {Hamer}, \citenamefont {Terry}, \citenamefont {Nayak},
  \citenamefont {Ahn}, \citenamefont {Pak}, \citenamefont {Lee}, \citenamefont
  {Sohn}, \citenamefont {Molas}, \citenamefont {Koperski}, \citenamefont
  {Watanabe}, \citenamefont {Taniguchi}, \citenamefont {Novoselov},
  \citenamefont {Gorbachev}, \citenamefont {Shin}, \citenamefont {Fal’ko},\
  and\ \citenamefont {Tartakovskii}}]{Alexeev_2019}%
  \BibitemOpen
  \bibfield  {author} {\bibinfo {author} {\bibfnamefont {E.~M.}\ \bibnamefont
  {Alexeev}}, \bibinfo {author} {\bibfnamefont {D.~A.}\ \bibnamefont
  {Ruiz-Tijerina}}, \bibinfo {author} {\bibfnamefont {M.}~\bibnamefont
  {Danovich}}, \bibinfo {author} {\bibfnamefont {M.~J.}\ \bibnamefont {Hamer}},
  \bibinfo {author} {\bibfnamefont {D.~J.}\ \bibnamefont {Terry}}, \bibinfo
  {author} {\bibfnamefont {P.~K.}\ \bibnamefont {Nayak}}, \bibinfo {author}
  {\bibfnamefont {S.}~\bibnamefont {Ahn}}, \bibinfo {author} {\bibfnamefont
  {S.}~\bibnamefont {Pak}}, \bibinfo {author} {\bibfnamefont {J.}~\bibnamefont
  {Lee}}, \bibinfo {author} {\bibfnamefont {J.~I.}\ \bibnamefont {Sohn}},
  \bibinfo {author} {\bibfnamefont {M.~R.}\ \bibnamefont {Molas}}, \bibinfo
  {author} {\bibfnamefont {M.}~\bibnamefont {Koperski}}, \bibinfo {author}
  {\bibfnamefont {K.}~\bibnamefont {Watanabe}}, \bibinfo {author}
  {\bibfnamefont {T.}~\bibnamefont {Taniguchi}}, \bibinfo {author}
  {\bibfnamefont {K.~S.}\ \bibnamefont {Novoselov}}, \bibinfo {author}
  {\bibfnamefont {R.~V.}\ \bibnamefont {Gorbachev}}, \bibinfo {author}
  {\bibfnamefont {H.~S.}\ \bibnamefont {Shin}}, \bibinfo {author}
  {\bibfnamefont {V.~I.}\ \bibnamefont {Fal’ko}},\ and\ \bibinfo {author}
  {\bibfnamefont {A.~I.}\ \bibnamefont {Tartakovskii}},\ }\bibfield  {title}
  {\bibinfo {title} {Resonantly hybridized excitons in moiré superlattices in
  van der {W}aals heterostructures},\ }\href
  {https://doi.org/10.1038/s41586-019-0986-9} {\bibfield  {journal} {\bibinfo
  {journal} {Nature}\ }\textbf {\bibinfo {volume} {567}},\ \bibinfo {pages}
  {81} (\bibinfo {year} {2019})}\BibitemShut {NoStop}%
\bibitem [{\citenamefont {Tang}\ \emph {et~al.}(2021)\citenamefont {Tang},
  \citenamefont {Gu}, \citenamefont {Liu}, \citenamefont {Watanabe},
  \citenamefont {Taniguchi}, \citenamefont {Hone}, \citenamefont {Mak},\ and\
  \citenamefont {Shan}}]{Tang_2021}%
  \BibitemOpen
  \bibfield  {author} {\bibinfo {author} {\bibfnamefont {Y.}~\bibnamefont
  {Tang}}, \bibinfo {author} {\bibfnamefont {J.}~\bibnamefont {Gu}}, \bibinfo
  {author} {\bibfnamefont {S.}~\bibnamefont {Liu}}, \bibinfo {author}
  {\bibfnamefont {K.}~\bibnamefont {Watanabe}}, \bibinfo {author}
  {\bibfnamefont {T.}~\bibnamefont {Taniguchi}}, \bibinfo {author}
  {\bibfnamefont {J.}~\bibnamefont {Hone}}, \bibinfo {author} {\bibfnamefont
  {K.}~\bibnamefont {Mak}},\ and\ \bibinfo {author} {\bibfnamefont
  {J.}~\bibnamefont {Shan}},\ }\bibfield  {title} {\bibinfo {title} {{Tuning
  layer-hybridized moir\'e excitons by the quantum-confined Stark effect}},\
  }\href {https://doi.org/10.1038/s41565-020-00783-2} {\bibfield  {journal}
  {\bibinfo  {journal} {Nat. Nanotechnol.}\ }\textbf {\bibinfo {volume} {16}},\
  \bibinfo {pages} {52} (\bibinfo {year} {2021})}\BibitemShut {NoStop}%
\bibitem [{\citenamefont {Tang}\ \emph {et~al.}(2022)\citenamefont {Tang},
  \citenamefont {Gu}, \citenamefont {Liu}, \citenamefont {Watanabe},
  \citenamefont {Taniguchi}, \citenamefont {Hone}, \citenamefont {Mak},\ and\
  \citenamefont {Shan}}]{Tang_2022}%
  \BibitemOpen
  \bibfield  {author} {\bibinfo {author} {\bibfnamefont {Y.}~\bibnamefont
  {Tang}}, \bibinfo {author} {\bibfnamefont {J.}~\bibnamefont {Gu}}, \bibinfo
  {author} {\bibfnamefont {S.}~\bibnamefont {Liu}}, \bibinfo {author}
  {\bibfnamefont {K.}~\bibnamefont {Watanabe}}, \bibinfo {author}
  {\bibfnamefont {T.}~\bibnamefont {Taniguchi}}, \bibinfo {author}
  {\bibfnamefont {J.~C.}\ \bibnamefont {Hone}}, \bibinfo {author}
  {\bibfnamefont {K.~F.}\ \bibnamefont {Mak}},\ and\ \bibinfo {author}
  {\bibfnamefont {J.}~\bibnamefont {Shan}},\ }\bibfield  {title} {\bibinfo
  {title} {{Dielectric catastrophe at the Wigner-Mott transition in a
  moir{\'{e}} superlattice}},\ }\href
  {https://doi.org/10.1038%2Fs41467-022-32037-1} {\bibfield  {journal}
  {\bibinfo  {journal} {Nat. Commun.}\ }\textbf {\bibinfo {volume} {13}},\
  \bibinfo {pages} {4271} (\bibinfo {year} {2022})}\BibitemShut {NoStop}%
\bibitem [{\citenamefont {Polovnikov}\ \emph {et~al.}(2022)\citenamefont
  {Polovnikov}, \citenamefont {Scherzer}, \citenamefont {Misra}, \citenamefont
  {Huang}, \citenamefont {Mohl}, \citenamefont {Li}, \citenamefont {Göser},
  \citenamefont {Förste}, \citenamefont {Bilgin},\ and\ \citenamefont
  {Watanabe}}]{Polovnikov2022}%
  \BibitemOpen
  \bibfield  {author} {\bibinfo {author} {\bibfnamefont {B.}~\bibnamefont
  {Polovnikov}}, \bibinfo {author} {\bibfnamefont {J.}~\bibnamefont
  {Scherzer}}, \bibinfo {author} {\bibfnamefont {S.}~\bibnamefont {Misra}},
  \bibinfo {author} {\bibfnamefont {X.}~\bibnamefont {Huang}}, \bibinfo
  {author} {\bibfnamefont {C.}~\bibnamefont {Mohl}}, \bibinfo {author}
  {\bibfnamefont {Z.}~\bibnamefont {Li}}, \bibinfo {author} {\bibfnamefont
  {J.}~\bibnamefont {Göser}}, \bibinfo {author} {\bibfnamefont
  {J.}~\bibnamefont {Förste}}, \bibinfo {author} {\bibfnamefont
  {I.}~\bibnamefont {Bilgin}},\ and\ \bibinfo {author} {\bibfnamefont
  {K.}~\bibnamefont {Watanabe}},\ }\bibfield  {title} {\bibinfo {title}
  {{Coulomb-correlated states of moir\'e excitons and elementary charges on a
  semiconductor moir\'e lattice at integer and fractional fillings}},\
  }\href@noop {} {\bibfield  {journal} {\bibinfo  {journal} {arXiv preprint
  arXiv:2208.04056}\ } (\bibinfo {year} {2022})}\BibitemShut {NoStop}%
\bibitem [{\citenamefont {Ross}\ \emph {et~al.}(2013)\citenamefont {Ross},
  \citenamefont {Wu}, \citenamefont {Yu}, \citenamefont {Ghimire},
  \citenamefont {Jones}, \citenamefont {Aivazian}, \citenamefont {Yan},
  \citenamefont {Mandrus}, \citenamefont {Xiao}, \citenamefont {Yao},\ and\
  \citenamefont {Xu}}]{Ross_2013}%
  \BibitemOpen
  \bibfield  {author} {\bibinfo {author} {\bibfnamefont {J.}~\bibnamefont
  {Ross}}, \bibinfo {author} {\bibfnamefont {S.}~\bibnamefont {Wu}}, \bibinfo
  {author} {\bibfnamefont {H.}~\bibnamefont {Yu}}, \bibinfo {author}
  {\bibfnamefont {N.}~\bibnamefont {Ghimire}}, \bibinfo {author} {\bibfnamefont
  {A.}~\bibnamefont {Jones}}, \bibinfo {author} {\bibfnamefont
  {G.}~\bibnamefont {Aivazian}}, \bibinfo {author} {\bibfnamefont {J.-Q.}\
  \bibnamefont {Yan}}, \bibinfo {author} {\bibfnamefont {D.}~\bibnamefont
  {Mandrus}}, \bibinfo {author} {\bibfnamefont {D.}~\bibnamefont {Xiao}},
  \bibinfo {author} {\bibfnamefont {W.}~\bibnamefont {Yao}},\ and\ \bibinfo
  {author} {\bibfnamefont {X.}~\bibnamefont {Xu}},\ }\bibfield  {title}
  {\bibinfo {title} {Electrical control of neutral and charged excitons in a
  monolayer semiconductor},\ }\href {https://doi.org/10.1038/ncomms2498}
  {\bibfield  {journal} {\bibinfo  {journal} {Nat. Commun.}\ }\textbf {\bibinfo
  {volume} {4}},\ \bibinfo {pages} {1474} (\bibinfo {year} {2013})}\BibitemShut
  {NoStop}%
\bibitem [{\citenamefont {Li}\ \emph {et~al.}(2014)\citenamefont {Li},
  \citenamefont {Ludwig}, \citenamefont {Low}, \citenamefont {Chernikov},
  \citenamefont {Cui}, \citenamefont {Arefe}, \citenamefont {Kim},
  \citenamefont {van~der Zande}, \citenamefont {Rigosi}, \citenamefont {Hill},
  \citenamefont {Kim}, \citenamefont {Hone}, \citenamefont {Li}, \citenamefont
  {Smirnov},\ and\ \citenamefont {Heinz}}]{Li2014}%
  \BibitemOpen
  \bibfield  {author} {\bibinfo {author} {\bibfnamefont {Y.}~\bibnamefont
  {Li}}, \bibinfo {author} {\bibfnamefont {J.}~\bibnamefont {Ludwig}}, \bibinfo
  {author} {\bibfnamefont {T.}~\bibnamefont {Low}}, \bibinfo {author}
  {\bibfnamefont {A.}~\bibnamefont {Chernikov}}, \bibinfo {author}
  {\bibfnamefont {X.}~\bibnamefont {Cui}}, \bibinfo {author} {\bibfnamefont
  {G.}~\bibnamefont {Arefe}}, \bibinfo {author} {\bibfnamefont {Y.~D.}\
  \bibnamefont {Kim}}, \bibinfo {author} {\bibfnamefont {A.~M.}\ \bibnamefont
  {van~der Zande}}, \bibinfo {author} {\bibfnamefont {A.}~\bibnamefont
  {Rigosi}}, \bibinfo {author} {\bibfnamefont {H.~M.}\ \bibnamefont {Hill}},
  \bibinfo {author} {\bibfnamefont {S.~H.}\ \bibnamefont {Kim}}, \bibinfo
  {author} {\bibfnamefont {J.}~\bibnamefont {Hone}}, \bibinfo {author}
  {\bibfnamefont {Z.}~\bibnamefont {Li}}, \bibinfo {author} {\bibfnamefont
  {D.}~\bibnamefont {Smirnov}},\ and\ \bibinfo {author} {\bibfnamefont {T.~F.}\
  \bibnamefont {Heinz}},\ }\bibfield  {title} {\bibinfo {title} {{Valley
  Splitting and Polarization by the Zeeman Effect in Monolayer
  ${\mathrm{MoSe}}_{2}$}},\ }\href
  {https://doi.org/10.1103/PhysRevLett.113.266804} {\bibfield  {journal}
  {\bibinfo  {journal} {Phys. Rev. Lett.}\ }\textbf {\bibinfo {volume} {113}},\
  \bibinfo {pages} {266804} (\bibinfo {year} {2014})}\BibitemShut {NoStop}%
\bibitem [{\citenamefont {MacNeill}\ \emph {et~al.}(2015)\citenamefont
  {MacNeill}, \citenamefont {Heikes}, \citenamefont {Mak}, \citenamefont
  {Anderson}, \citenamefont {Kormányos}, \citenamefont {Zólyomi},
  \citenamefont {Park},\ and\ \citenamefont {Ralph}}]{MacNeill2015}%
  \BibitemOpen
  \bibfield  {author} {\bibinfo {author} {\bibfnamefont {D.}~\bibnamefont
  {MacNeill}}, \bibinfo {author} {\bibfnamefont {C.}~\bibnamefont {Heikes}},
  \bibinfo {author} {\bibfnamefont {K.~F.}\ \bibnamefont {Mak}}, \bibinfo
  {author} {\bibfnamefont {Z.}~\bibnamefont {Anderson}}, \bibinfo {author}
  {\bibfnamefont {A.}~\bibnamefont {Kormányos}}, \bibinfo {author}
  {\bibfnamefont {V.}~\bibnamefont {Zólyomi}}, \bibinfo {author}
  {\bibfnamefont {J.}~\bibnamefont {Park}},\ and\ \bibinfo {author}
  {\bibfnamefont {D.~C.}\ \bibnamefont {Ralph}},\ }\bibfield  {title} {\bibinfo
  {title} {{Breaking of Valley Degeneracy by Magnetic Field in Monolayer
  ${\mathrm{MoSe}}_{2}$}},\ }\href
  {https://doi.org/10.1103/PhysRevLett.114.037401} {\bibfield  {journal}
  {\bibinfo  {journal} {Phys. Rev. Lett.}\ }\textbf {\bibinfo {volume} {114}},\
  \bibinfo {pages} {037401} (\bibinfo {year} {2015})}\BibitemShut {NoStop}%
\bibitem [{\citenamefont {Koperski}\ \emph {et~al.}(2019)\citenamefont
  {Koperski}, \citenamefont {Molas}, \citenamefont {Arora}, \citenamefont
  {Nogajewski}, \citenamefont {Bartos}, \citenamefont {Wyzula}, \citenamefont
  {Vaclavkova}, \citenamefont {Kossacki},\ and\ \citenamefont
  {Potemski}}]{Koperski2019}%
  \BibitemOpen
  \bibfield  {author} {\bibinfo {author} {\bibfnamefont {M.}~\bibnamefont
  {Koperski}}, \bibinfo {author} {\bibfnamefont {M.~R.}\ \bibnamefont {Molas}},
  \bibinfo {author} {\bibfnamefont {A.}~\bibnamefont {Arora}}, \bibinfo
  {author} {\bibfnamefont {K.}~\bibnamefont {Nogajewski}}, \bibinfo {author}
  {\bibfnamefont {M.}~\bibnamefont {Bartos}}, \bibinfo {author} {\bibfnamefont
  {J.}~\bibnamefont {Wyzula}}, \bibinfo {author} {\bibfnamefont
  {D.}~\bibnamefont {Vaclavkova}}, \bibinfo {author} {\bibfnamefont
  {P.}~\bibnamefont {Kossacki}},\ and\ \bibinfo {author} {\bibfnamefont
  {M.}~\bibnamefont {Potemski}},\ }\bibfield  {title} {\bibinfo {title}
  {{Orbital, spin and valley contributions to Zeeman splitting of excitonic
  resonances in MoSe$_2$, WSe$_2$ and WS$_2$ Monolayers}},\ }\href
  {https://doi.org/10.1088/2053-1583/aae14b} {\bibfield  {journal} {\bibinfo
  {journal} {2D Mater.}\ }\textbf {\bibinfo {volume} {6}},\ \bibinfo {pages}
  {015001} (\bibinfo {year} {2019})}\BibitemShut {NoStop}%
\bibitem [{\citenamefont {Goryca}\ \emph {et~al.}(2019)\citenamefont {Goryca},
  \citenamefont {Li}, \citenamefont {Stier}, \citenamefont {Taniguchi},
  \citenamefont {Watanabe}, \citenamefont {Courtade}, \citenamefont {Shree},
  \citenamefont {Robert}, \citenamefont {Urbaszek}, \citenamefont {Marie},\
  and\ \citenamefont {Crooker}}]{Goryca2019}%
  \BibitemOpen
  \bibfield  {author} {\bibinfo {author} {\bibfnamefont {M.}~\bibnamefont
  {Goryca}}, \bibinfo {author} {\bibfnamefont {J.}~\bibnamefont {Li}}, \bibinfo
  {author} {\bibfnamefont {A.~V.}\ \bibnamefont {Stier}}, \bibinfo {author}
  {\bibfnamefont {T.}~\bibnamefont {Taniguchi}}, \bibinfo {author}
  {\bibfnamefont {K.}~\bibnamefont {Watanabe}}, \bibinfo {author}
  {\bibfnamefont {E.}~\bibnamefont {Courtade}}, \bibinfo {author}
  {\bibfnamefont {S.}~\bibnamefont {Shree}}, \bibinfo {author} {\bibfnamefont
  {C.}~\bibnamefont {Robert}}, \bibinfo {author} {\bibfnamefont
  {B.}~\bibnamefont {Urbaszek}}, \bibinfo {author} {\bibfnamefont
  {X.}~\bibnamefont {Marie}},\ and\ \bibinfo {author} {\bibfnamefont {S.~A.}\
  \bibnamefont {Crooker}},\ }\bibfield  {title} {\bibinfo {title} {Revealing
  exciton masses and dielectric properties of monolayer semiconductors with
  high magnetic fields},\ }\href {https://doi.org/10.1038/s41467-019-12180-y}
  {\bibfield  {journal} {\bibinfo  {journal} {Nat. Commun.}\ }\textbf {\bibinfo
  {volume} {10}},\ \bibinfo {pages} {4172} (\bibinfo {year}
  {2019})}\BibitemShut {NoStop}%
\bibitem [{\citenamefont {Lundt}\ \emph {et~al.}(2019)\citenamefont {Lundt},
  \citenamefont {Klaas}, \citenamefont {Sedov}, \citenamefont {Waldherr},
  \citenamefont {Knopf}, \citenamefont {Blei}, \citenamefont {Tongay},
  \citenamefont {Klembt}, \citenamefont {Taniguchi}, \citenamefont {Watanabe},
  \citenamefont {Schulz}, \citenamefont {Kavokin}, \citenamefont {H\"ofling},
  \citenamefont {Eilenberger},\ and\ \citenamefont {Schneider}}]{Lundt_2019}%
  \BibitemOpen
  \bibfield  {author} {\bibinfo {author} {\bibfnamefont {N.}~\bibnamefont
  {Lundt}}, \bibinfo {author} {\bibfnamefont {M.}~\bibnamefont {Klaas}},
  \bibinfo {author} {\bibfnamefont {E.}~\bibnamefont {Sedov}}, \bibinfo
  {author} {\bibfnamefont {M.}~\bibnamefont {Waldherr}}, \bibinfo {author}
  {\bibfnamefont {H.}~\bibnamefont {Knopf}}, \bibinfo {author} {\bibfnamefont
  {M.}~\bibnamefont {Blei}}, \bibinfo {author} {\bibfnamefont {S.}~\bibnamefont
  {Tongay}}, \bibinfo {author} {\bibfnamefont {S.}~\bibnamefont {Klembt}},
  \bibinfo {author} {\bibfnamefont {T.}~\bibnamefont {Taniguchi}}, \bibinfo
  {author} {\bibfnamefont {K.}~\bibnamefont {Watanabe}}, \bibinfo {author}
  {\bibfnamefont {U.}~\bibnamefont {Schulz}}, \bibinfo {author} {\bibfnamefont
  {A.}~\bibnamefont {Kavokin}}, \bibinfo {author} {\bibfnamefont
  {S.}~\bibnamefont {H\"ofling}}, \bibinfo {author} {\bibfnamefont
  {F.}~\bibnamefont {Eilenberger}},\ and\ \bibinfo {author} {\bibfnamefont
  {C.}~\bibnamefont {Schneider}},\ }\bibfield  {title} {\bibinfo {title}
  {Magnetic-field-induced splitting and polarization of monolayer-based valley
  exciton polaritons},\ }\href {https://doi.org/10.1103/PhysRevB.100.121303}
  {\bibfield  {journal} {\bibinfo  {journal} {Phys. Rev. B}\ }\textbf {\bibinfo
  {volume} {100}},\ \bibinfo {pages} {121303} (\bibinfo {year}
  {2019})}\BibitemShut {NoStop}%
\bibitem [{\citenamefont {Ma}\ \emph {et~al.}(2022)\citenamefont {Ma},
  \citenamefont {Wang}, \citenamefont {Chung}, \citenamefont {Pfeiffer},
  \citenamefont {West}, \citenamefont {Baldwin}, \citenamefont {Winkler},\ and\
  \citenamefont {Shayegan}}]{Ma_2022}%
  \BibitemOpen
  \bibfield  {author} {\bibinfo {author} {\bibfnamefont {M.~K.}\ \bibnamefont
  {Ma}}, \bibinfo {author} {\bibfnamefont {C.}~\bibnamefont {Wang}}, \bibinfo
  {author} {\bibfnamefont {Y.~J.}\ \bibnamefont {Chung}}, \bibinfo {author}
  {\bibfnamefont {L.~N.}\ \bibnamefont {Pfeiffer}}, \bibinfo {author}
  {\bibfnamefont {K.~W.}\ \bibnamefont {West}}, \bibinfo {author}
  {\bibfnamefont {K.~W.}\ \bibnamefont {Baldwin}}, \bibinfo {author}
  {\bibfnamefont {R.}~\bibnamefont {Winkler}},\ and\ \bibinfo {author}
  {\bibfnamefont {M.}~\bibnamefont {Shayegan}},\ }\bibfield  {title} {\bibinfo
  {title} {{Robust Quantum Hall Ferromagnetism near a Gate-Tuned
  $\ensuremath{\nu}=1$ Landau Level Crossing}},\ }\href
  {https://doi.org/10.1103/PhysRevLett.129.196801} {\bibfield  {journal}
  {\bibinfo  {journal} {Phys. Rev. Lett.}\ }\textbf {\bibinfo {volume} {129}},\
  \bibinfo {pages} {196801} (\bibinfo {year} {2022})}\BibitemShut {NoStop}%
\bibitem [{\citenamefont {Liu}\ \emph {et~al.}(2022)\citenamefont {Liu},
  \citenamefont {Farahi}, \citenamefont {Chiu}, \citenamefont {Papic},
  \citenamefont {Watanabe}, \citenamefont {Taniguchi}, \citenamefont
  {Zaletel},\ and\ \citenamefont {Yazdani}}]{Liu_2022}%
  \BibitemOpen
  \bibfield  {author} {\bibinfo {author} {\bibfnamefont {X.}~\bibnamefont
  {Liu}}, \bibinfo {author} {\bibfnamefont {G.}~\bibnamefont {Farahi}},
  \bibinfo {author} {\bibfnamefont {C.-L.}\ \bibnamefont {Chiu}}, \bibinfo
  {author} {\bibfnamefont {Z.}~\bibnamefont {Papic}}, \bibinfo {author}
  {\bibfnamefont {K.}~\bibnamefont {Watanabe}}, \bibinfo {author}
  {\bibfnamefont {T.}~\bibnamefont {Taniguchi}}, \bibinfo {author}
  {\bibfnamefont {M.~P.}\ \bibnamefont {Zaletel}},\ and\ \bibinfo {author}
  {\bibfnamefont {A.}~\bibnamefont {Yazdani}},\ }\bibfield  {title} {\bibinfo
  {title} {{Visualizing broken symmetry and topological defects in a quantum
  Hall ferromagnet}},\ }\href
  {https://www.science.org/doi/abs/10.1126/science.abm3770} {\bibfield
  {journal} {\bibinfo  {journal} {Science}\ }\textbf {\bibinfo {volume}
  {375}},\ \bibinfo {pages} {321} (\bibinfo {year} {2022})}\BibitemShut
  {NoStop}%
\bibitem [{\citenamefont {Ravets}\ \emph {et~al.}(2018)\citenamefont {Ravets},
  \citenamefont {Kn\"uppel}, \citenamefont {Faelt}, \citenamefont {Cotlet},
  \citenamefont {Kroner}, \citenamefont {Wegscheider},\ and\ \citenamefont
  {Imamoglu}}]{Ravets_2018}%
  \BibitemOpen
  \bibfield  {author} {\bibinfo {author} {\bibfnamefont {S.}~\bibnamefont
  {Ravets}}, \bibinfo {author} {\bibfnamefont {P.}~\bibnamefont {Kn\"uppel}},
  \bibinfo {author} {\bibfnamefont {S.}~\bibnamefont {Faelt}}, \bibinfo
  {author} {\bibfnamefont {O.}~\bibnamefont {Cotlet}}, \bibinfo {author}
  {\bibfnamefont {M.}~\bibnamefont {Kroner}}, \bibinfo {author} {\bibfnamefont
  {W.}~\bibnamefont {Wegscheider}},\ and\ \bibinfo {author} {\bibfnamefont
  {A.}~\bibnamefont {Imamoglu}},\ }\bibfield  {title} {\bibinfo {title}
  {{Polaron Polaritons in the Integer and Fractional Quantum Hall Regimes}},\
  }\href {https://doi.org/10.1103/PhysRevLett.120.057401} {\bibfield  {journal}
  {\bibinfo  {journal} {Phys. Rev. Lett.}\ }\textbf {\bibinfo {volume} {120}},\
  \bibinfo {pages} {057401} (\bibinfo {year} {2018})}\BibitemShut {NoStop}%
\bibitem [{\citenamefont {Kn\"uppel}\ \emph {et~al.}(2019)\citenamefont
  {Kn\"uppel}, \citenamefont {Ravets}, \citenamefont {Kroner}, \citenamefont
  {F\"alt}, \citenamefont {Wegscheider},\ and\ \citenamefont
  {Imamoglu}}]{Knueppel_2019}%
  \BibitemOpen
  \bibfield  {author} {\bibinfo {author} {\bibfnamefont {P.}~\bibnamefont
  {Kn\"uppel}}, \bibinfo {author} {\bibfnamefont {S.}~\bibnamefont {Ravets}},
  \bibinfo {author} {\bibfnamefont {M.}~\bibnamefont {Kroner}}, \bibinfo
  {author} {\bibfnamefont {S.}~\bibnamefont {F\"alt}}, \bibinfo {author}
  {\bibfnamefont {W.}~\bibnamefont {Wegscheider}},\ and\ \bibinfo {author}
  {\bibfnamefont {A.}~\bibnamefont {Imamoglu}},\ }\bibfield  {title} {\bibinfo
  {title} {{Nonlinear optics in the fractional quantum Hall regime}},\ }\href
  {https://doi.org/10.1038/s41586-019-1356-3} {\bibfield  {journal} {\bibinfo
  {journal} {Nature}\ }\textbf {\bibinfo {volume} {572}},\ \bibinfo {pages}
  {91} (\bibinfo {year} {2019})}\BibitemShut {NoStop}%
\bibitem [{\citenamefont {Leggett}(2006)}]{Leggett2006}%
  \BibitemOpen
  \bibfield  {author} {\bibinfo {author} {\bibfnamefont {A.~J.}\ \bibnamefont
  {Leggett}},\ }\href@noop {} {\emph {\bibinfo {title} {{Quantum liquids: Bose
  condensation and Cooper pairing in condensed-matter systems}}}}\ (\bibinfo
  {publisher} {Oxford university press},\ \bibinfo {year} {2006})\BibitemShut
  {NoStop}%
\end{thebibliography}
%

\end{document}


\title{Correlated magnetism of moir\'e exciton-polaritons on a triangular electron-spin lattice \\ \vspace{7pt} \Large{Supplementary Information}}

\author{Johannes Scherzer$^{1,*}$, Lukas Lackner$^{2,*}$, Bo Han$^{2}$, Borislav Polovnikov$^{1}$ , Lukas Husel$^{1}$, Jonas Göser$^{1}$, Zhijie Li$^{1}$, Jens-Christian Drawer$^{2}$, Martin Esmann$^{2}$, Christoph Bennenhei$^{2}$, Falk Eilenberger$^{3,4}$, Kenji Watanabe$^{5}$, Takashi Taniguchi$^{6}$, Anvar S. Baimuratov$^{1}$, Christian Schneider$^{2}$, Alexander H\"ogele$^{1,7}$ \footnote[0]{$^{*}$ These authors contributed equally to this work.}}


    \affiliation{$^1$Fakult\"at f\"ur Physik, Munich Quantum Center, and Center for NanoScience (CeNS), Ludwig-Maximilians-Universit\"at M\"unchen, Geschwister-Scholl-Platz 1, 80539 M\"unchen, Germany}

    \affiliation{$^2$Institute of Physics, Carl von Ossietzky Universit\"at Oldenburg, Carl-von-Ossietzky-Straße 9-11, 26129 Oldenburg, Germany}
    
    \affiliation{$^3$Institute of Applied Physics, Friedrich-Schiller-Universit\"at Jena, Max-Wien-Platz 1, 07743 Jena, Germany}    
    
    \affiliation{$^4$Fraunhofer-Institute of Applied Optics and Precision Engineering IOF, Albert-Einstein-Straße 7, 07745 Jena, Germany}
    
    \affiliation{$^5$Research Center for Electronic and Optical Materials, National Institute for Materials Science, 1-1 Namiki, Tsukuba 305-0044, Japan}
    
    \affiliation{$^6$Research Center for Materials Nanoarchitectonics, National Institute for Materials Science,  1-1 Namiki, Tsukuba 305-0044, Japan}
       
    \affiliation{$^7$Munich Center for Quantum Science and Technology (MCQST), Schellingstra\ss{}e 4, 80799 M\"unchen, Germany}

\maketitle

\noindent {\textbf{\large{Supplementary Note 1: Characterization of the MoSe$_{2}$/WS$_{2}$ heterostructure}}\\

The dual-gate layout of the field-effect device with symmetric hBN layer thickness on the top and bottom of the MoSe$_{2}$/WS$_{2}$ heterobilayer (HBL) allows to subject the moir\'e excitons to perpendicular electric field $F$, or to vary the doping level V$_{\mu}$. The former is achieved by an imbalanced tuning of the top and bottom gate voltages V$_{\text{T}}$ and V$_{\text{B}}$ as $\Delta V_{\text{TB}} = (V_{\text{B}} - V_{\text{T}}) = Fd_{\text{hBN}} $ with respect to the grounded reservoir in contact with both MoSe$_{2}$ and WS$_{2}$ monolayers ($d_{\text{hBN}}$ denotes the total hBN thickness between the top and bottom graphite gates) and probes the static out-of-plane exciton dipole moment via the Stark effect. The latter is achieved by balancing both gates and tuning them simultaneously against the ground as V$_{\text{G}} = (V_{\text{T}} + V_{\text{B}})/2$ to shift the Fermi level through the spatially modulated electron and hole potentials of the moir\'e heterostructure. 

We first focus on the charge-neutral regime of the MoSe$_2$/WS$_2$ heterostack. The top panel of Fig.~\ref{SMfig1}a shows a cryogenic differential reflection spectrum $\textrm{DR} = (R_0-R)/R_0$, with $R$ being the reflection spectrum at the spot of interest and $R_0$ a reference background away from the heterostructure, exhibiting two bright moir\'e exciton resonances, M$_1$ and M$_2$ at $1.587$ and $1.622$~eV, respectively. The corresponding photoluminescence (PL) spectrum in the lower panel of Fig.~\ref{SMfig1}a indicates a type-I band alignment, with only one emission peak of the ground state moir\'e exciton M$_1$ due to efficient population relaxation from higher-energy moir\'e exciton states. As we vary the out-of-plane electric field, we observe a vanishingly small dispersion of M$_1$ and M$_2$ in  Fig.~\ref{SMfig1}b consistent with their predominant MoSe$_{2}$ intralayer exciton character. Using the detailed analysis of the moir\'e exciton band structure in Ref.~\cite{Polovnikov_2023} and the fitting parameters of the moir\'e
exciton potential $V_\text{moir\'e}$ shown in Fig.~1b of the main text, we obtained the spatial distributions of M$_1$ and M$_2$ moir\'e exciton wavefunctions as shown in Fig.~\ref{SMfig1}c.

\begin{figure*}[htb]
\includegraphics[scale=1.1]{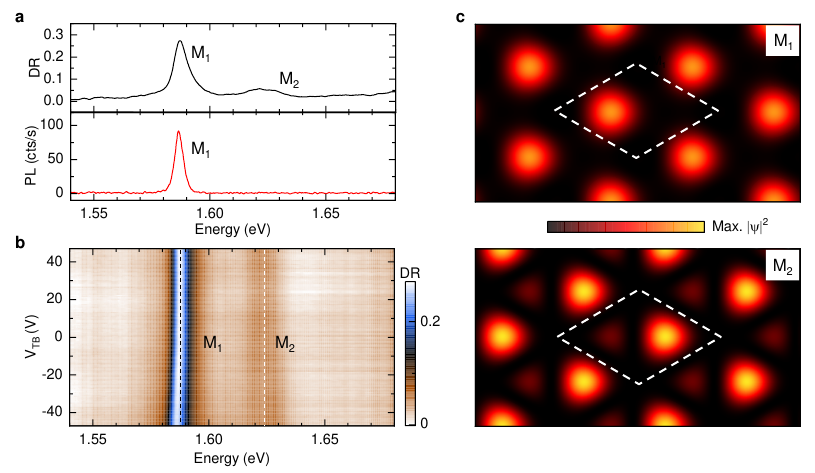}
\vspace{-8pt}
\caption{\textbf{Neutral moir\'e excitons.} \textbf{a}, Top panel: Charge neutral reflection contrast spectrum without top DBR in the region of interest on the heterobilayer. Bottom panel: Photoluminescence spectrum at the same position. \textbf{b}, DR response of charge neutral moir\'e excitons M$_{1}$ and M$_{2}$ subjected to an out-of-plane electric field via the gate voltage $\Delta V_{\text{TB}}$. \textbf{c}, Amplitude distribution of the exciton wavefunctions within the moir\'e unit cell (white dashed lines) for M$_{1}$ (upper panel) and M$_{2}$ (lower panel) moir\'e excitons.}
\label{SMfig1}
\end{figure*}

Next, we study the optical response of the system in the regime of electron doping. Shifting the Fermi level by a positive imbalanced increase of V$_{\text{B}}$ and V$_{\text{T}}$ leads to the evolution of charged moir\'e excitons M$_{1}^{-}$ and M$_{2}^{-}$, as shown in Fig.~1c of the main text. To derive the electron density in the heterostructure at a specific gate voltage in units of charges per moir\'e unit cell, we inspect the optical response (minimum of the derivative $d(\text{DR})/dE$ between $1.54$ and $1.65$~eV) as a function of V$_{\text{B}}$ and V$_{\text{T}}$ in Fig.~\ref{SMfig_capacitor}a in the framework of the quantum capacitor model \cite{Popert_2022}.
Lines of constant color in Fig.~\ref{SMfig_capacitor}a correspond to constant optical response and thus constant electron density in the MoSe$_{2}$ layer. The latter derives from the geometric capacitance of the device and the moir\'e charge density $n_0 \approx 2.7 \times 10^{12}$~cm$^{-2}$, which is determined by the relative twist angle between the two layers ($1-2^\circ$). The conduction band offset $\Delta _{\text{CB}}=E_{\text{CB}}^W - E_{\text{CB}}^M$ between the WS$_2$ and MoSe$_2$ conduction bands and the layer specific density of states dependent on the Fermi level (see Fig.~\ref{SMfig_capacitor}d) are used as free parameters in the capacitor model simulations. The geometric capacitance and moir\'e density are fixed by the sample geometry, and the dielectric constants were set to $\epsilon _{\text{hBN}}= 4$ and $\epsilon _{\text{TMD}}= 8$. The conduction band offset $\Delta _{\text{CB}}=55$~meV derived from simulations is consistent with the expected type-I band alignment of the heterostructure. Fig.~\ref{SMfig_capacitor}c shows the results of the simulation for the electron density in the MoSe$_2$ layer in units of electrons per moir\'e unit cell as a function of V$_{\text{B}}$ and V$_{\text{T}}$. Lines of constant integer filling ($\nu=1,2,..$) agree with the main specific features of electron doping in the optical response in Fig.~\ref{SMfig_capacitor}a. The relevant transition from the charge neutral to the electron-doping regime, in particular at electron fillings $\nu=0 $ and $\nu=1 $, is marked by the dashed squares in Fig.~\ref{SMfig_capacitor}a, and the corresponding DR spectra are shown in Fig.~\ref{SMfig_capacitor}b. We note that for cavity experiments reported in the main text, the gate voltage was applied to the bottom gate V$_{\text{G}}=V_{\text{B}}$ only (V$_{\text{T}}=0$~V) for technical reasons.

\begin{figure*}[htb]  
\includegraphics[scale=1.2]{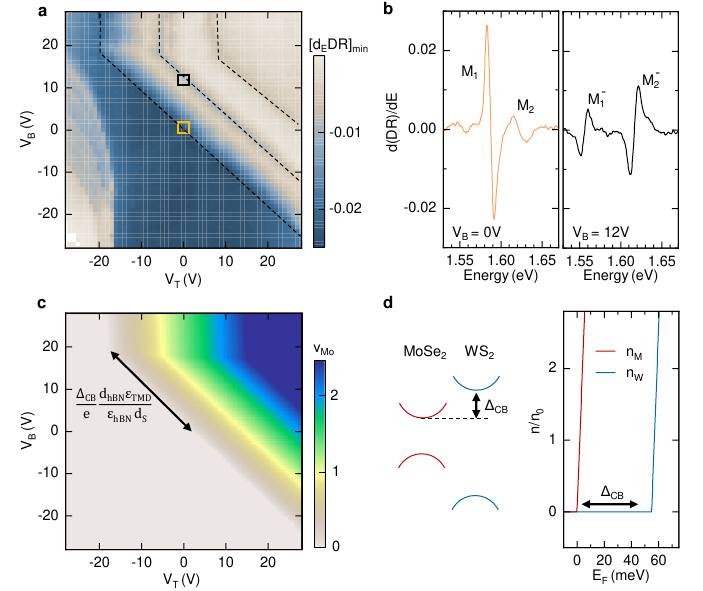}
\vspace{-8pt}
\caption{\textbf{Capacitor model analysis in the regime of electron doping.} \textbf{a}, Minimum of the derivative $d(\text{DR})/dE$ in the range between $1.55$ and $1.65$~eV as a function of V$_{\text{T}}$ and V$_{\text{B}}$. The black and yellow squares mark the gate voltage configuration at electron filling $\nu = 0$ (V$_{\text{B}} = V_{\text{T}} = 0$~V) and $\nu = 1$ (V$_{\text{B}} \approx 12$~V, V$_{\text{T}}=0$~V) discussed in the main text. The corresponding spectra are shown in \textbf{b}. \textbf{c}, Simulated electron density in the MoSe$_2$ layer in units of electrons per moir\'e unit cell as a function of V$_{\text{T}}$ and V$_{\text{B}}$, using dielectric constants of $\epsilon _{\text{hBN}}= 4$ and $\epsilon _{\text{TMD}}= 8$ for hBN and TMD layers, respectively ($d_S = 0.6$~nm denotes the distance between the two TMD layers, and $d_{\text{hBN}} = 180$~nm is the total hBN thickness between the top and bottom gates). \textbf{d}, Left panel: Schematic of the type-I band alignment in the MoSe$_2$/WS$_2$ heterostructure. Right panel: Electron density in the MoSe$_2$ (n$_\text{M}$) and WS$_2$ (n$_\text{W}$) layer as a function of the Fermi energy E$_{\text{F}}$ used in the capacitor model with conduction band offset $\Delta _{\text{CB}} = 55 $~meV.}
\label{SMfig_capacitor}
\end{figure*}

\clearpage
\noindent {\textbf{\large{Supplementary Note 2: Spectral analysis of moir\'e exciton, cavity and exciton-polariton resonances}}\\

\begin{figure*}[b!]  
\includegraphics[scale=1.07]{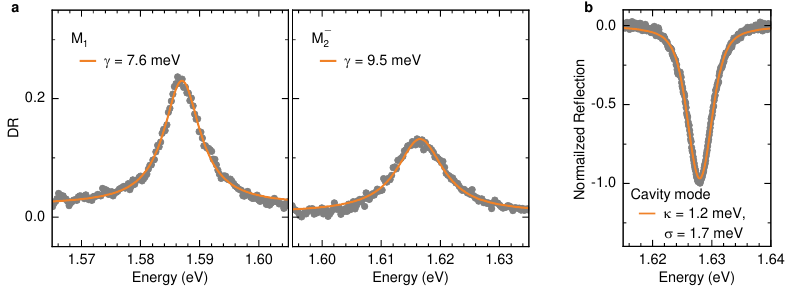}
\vspace{-8pt}
\caption{\textbf{Exciton and cavity resonances and linewidths.} \textbf{a}, Differential reflection spectra of M$_1$ (left panel) and M$_{2}^{-}$ (right panel) moir\'e exciton resonances with Lorentzian fits (orange solid lines) and FWHM broadening $\gamma$. \textbf{b}, Normalized reflection spectrum of the fundamental Laguerre-Gaussian cavity mode with a Voigt profile fit (orange solid line) with the homogeneous and inhomogeneous broadenings $\kappa$ and $\alpha$ of the respective Lorentzian and Gaussian contributions.} 
\label{SMfig_linewidths}
\end{figure*}

To determine the characteristic energies and spectral linewidths of the exciton-polariton system, we first analyzed the optical response of the moir\'e exciton absorption in confocal spectroscopy and the reflection contrast of the empty cavity. Fig.~\ref{SMfig_linewidths}a shows DR spectra of M$_{1}$ and M$_{2}^{-}$ moir\'e exciton resonances. The spectral lineshapes were fitted by Lorentzian functions 
\begin{equation}
L(E)=\dfrac{A}{\pi}\dfrac{\gamma /2}{E^2+(\gamma /2) ^2}
\end{equation}
with normalization coefficient $A$, yielding full-width at half-maximum (FWHM) linewidths $\gamma = 7.6$ and $9.5$~meV for the M$_{1}$ and M$_{2}^{-}$ resonance, respectively. 

In the back-scattering geometry of the microcavity reflection measurements, interference effects from reflections at different interfaces of the glass mesa with the DBR mirror on top modify the lineshape of the reflection spectra. To extract the symmetric lineshape of the fundamental Laguerre-Gaussian mode of the cavity we introduce a phenomenological phase factor $\beta$. The phase corrected reflection signal $\tilde{R}=Re^{i\beta}$ is subsequently corrected for a smooth background and normalized to facilitate the peak fitting procedure. The corrected spectral shape of the cavity mode, shown in Fig.~\ref{SMfig_linewidths}b, is inhomogenously broadened by cavity length fluctuations due to vibrations of the closed-cycle cryostat and is therefore best described by a Voigt profile as a convolution of a Gaussian and a Lorentzian function $G(E)$ and $L(E)$, respectively:
\begin{eqnarray}
V(E) & = & \int G(E')L(E-E')dE',\\
G(E) & = & \dfrac{1}{\sqrt{2 \pi} \sigma} e^{-E^2/(2 \sigma ^2)},
\end{eqnarray}
with the Gaussian broadening parameter $\sigma$. In Fig.~\ref{SMfig_linewidths}b, the FWHM of the Lorentzian yields the homogeneous cavity linewidth $\kappa = 1.2$~meV, and the FWHM of the Gaussian contribution $\sigma = 1.7$~meV accounts for the vibration-induced inhomogeneous broadening. As the vibrations occur in the frequency range between $1$~Hz and $1$~kHz, they are irrelevant on the timescale of coherent light-matter coupling.

\begin{figure*}[t!]  
\includegraphics[scale=1.07]{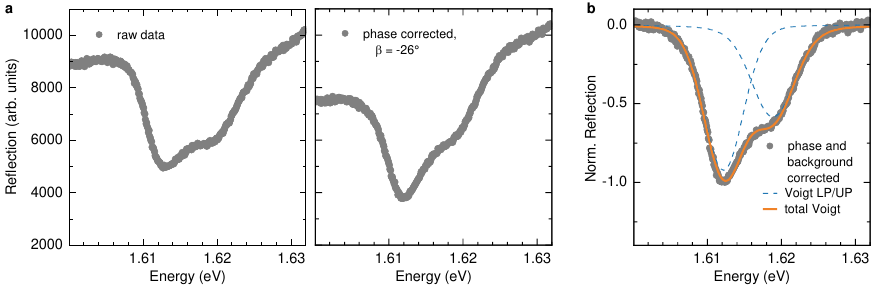}
\vspace{-8pt}
\caption{\textbf{Deconvolution of cavity reflection spectra.} \textbf{a}, Left panel: Raw reflection spectrum of the Zeeman-split M$_{2}^{-}$ polariton for a fixed cavity detuning and magnetic field. Right panel: Corresponding phase-corrected spectrum. \textbf{b}, Polynomial background correction and normalization of the phase-corrected spectrum allow to fit the lineshape with two Voigt profiles (orange solid line) of the corresponding upper (UP) and lower (LP) polariton resonances shown by blue dashed lines.}
\label{SMfig_spec_correction}
\end{figure*}

The reflection spectra of exciton-polaritons were analyzed accordingly from their respective raw spectra by applying both phase (with $\beta = -26^\circ$) and background correction, as shown in Fig.~\ref{SMfig_spec_correction}. The polariton peak energies were determined by fitting the corrected lineshape using a Voigt profile for each polariton branch:
\begin{equation}
V(E)=[G(L_1+L_2)](E)=\int G(E')L_1(E-E')+G(E')L_2(E-E')dE',
\end{equation}
as shown in Fig.~\ref{SMfig_spec_correction}b. Here, the contributions of two Lorentzians $L_1$ and $L_2$ account for the upper and lower polariton resonance, and the Gaussian function $G$ for vibration-induced broadening as well as for spatial inhomogeneity of the underlying exciton resonance within the waist of the cavity mode.

\noindent {\textbf{\large{Supplementary Note 3: Effective moir\'e polariton Land\'e factors}}\\

To model the evolution of the upper and lower polariton states with the magnetic field, we used a coupled oscillator model with a $4\times4$ Hamiltonian, with columns and rows related to the exciton transition with energy $E_\text{X}^{+}$ ($E_\text{X}^{-}$) and $\sigma^+$ ($\sigma^-$) polarization coupled to the cavity field with energy $E_C$ and $\sigma^+$ ($\sigma^-)$ polarization at Rabi splitting $\Omega^{+}$ ($\Omega^{-}$):
\begin{equation}
    H = 
    \begin{pmatrix}
        E_\text{X}^{+} & \Omega^{+}/2 & 0 & 0     \\
        \Omega^{+}/2 & E_\text{C} & 0 & 0 \\
        0 & 0 & E_\text{X}^{-} & \Omega^{-}/2     \\
        0 & 0 & \Omega^{-}/2 & E_\text{C}
    \end{pmatrix}.
    \label{Ham}
\end{equation} 

The cavity dispersion $E_{\text{C}}$ is assumed to be a linear function of the cavity length and independent of the external magnetic field. The eigenenergies of the Hamiltonian in Eq.~\eqref{Ham} corresponding to the upper and lower polariton branch UP and LP, respectively, with $\sigma^+$ and $\sigma^-$ polarization ($\eta = +/-$) are given by 
\begin{equation}
    E_{\text{UP/LP}}^\eta =
    \frac{E_\text{C} + E_\text{X}^\eta}{2} \pm \frac{1}{2} \sqrt{(E_\text{C}-E_\text{X}^\eta)^2+ {(\Omega ^\eta)^2}}.
\label{coupled_oscillator}
\end{equation}

\begin{figure*}[t!]  
\includegraphics[scale=1.1]{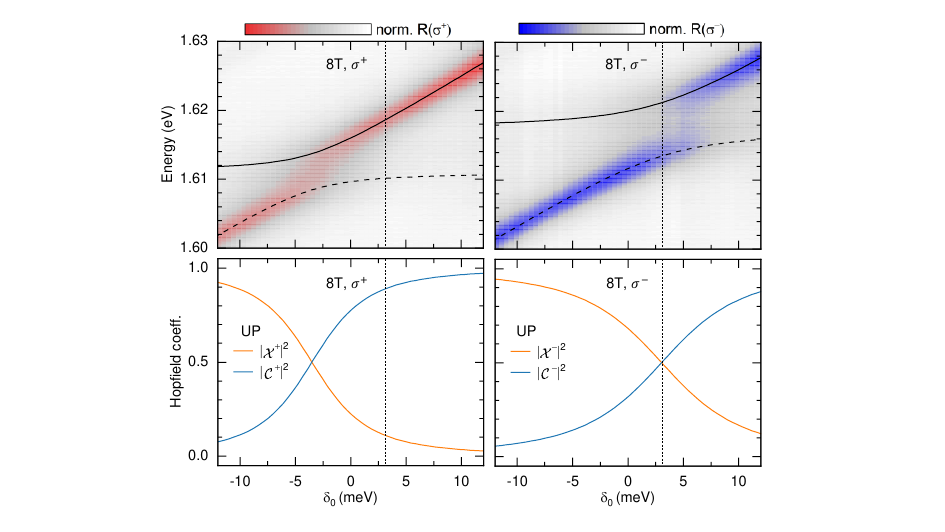}
\vspace{-40pt}
\caption{\textbf{Polarization-contrasting Hopfield coefficients of M$_{2}^{-}$ upper polariton branch in a magnetic field of $8$~T.} Upper panel: Dispersion of the reflection contrast as a function of the cavity length $\delta_{\text{0}}$ for the M$_{2}^{-}$ polariton at $\text{B}=8$~T for $\sigma^+$ (left panel) and $\sigma^-$ (right panel) polarization. The solutions of Eq.~\eqref{coupled_oscillator} for the upper and lower polariton branches are shown by solid and dashed lines, respectively. Lower panel: Corresponding exciton and photon Hopfield coefficients of the upper polariton branch with $\sigma ^+$ and $\sigma ^-$ polarization. The dashed vertical lines mark the cavity length for zero resonance detuning in $\sigma^-$ polarization, as discussed in Fig.~2 of the main text, where the exciton Hopfield coefficients differ significantly for the two circular polarizations ($|\mathcal{X}^+|^2 = 0.15$ for $\sigma ^+$ and $|\mathcal{X}^-|^2 = 0.50$ for $\sigma ^-$ polarization).} 
\label{SMfig_hopfield}
\end{figure*}

The upper panels of Fig.~\ref{SMfig_hopfield} show polarization-resolved cavity sweeps around the M$_{2}^{-}$ polaron resonance at a magnetic field of $8$~T, with model dispersions of the polariton branches as a function of the cavity detuning given by the eigenenergies in Eq.~\eqref{coupled_oscillator}. The Rabi splittings and exciton energies obtained from the fit were used to calculate the polarization-dependent exciton and photon Hopfield coefficients for a given cavity detuning and magnetic field as:
\begin{align}
|\mathcal{X}^\eta|^2=\dfrac{E_{\text{UP}}^{\eta}E_\text{X}^{\eta}-E_{\text{LP}}^{\eta}E_\text{C}}{(E_\text{C}+E_\text{X}^{\eta})\sqrt{(E_\text{C}-E_\text{X}^{\eta})^2+{(\Omega ^\eta)^2}}}\\    |\mathcal{C}^\eta|^2=\dfrac{E_{\text{UP}}^{\eta}E_\text{C}-E_{\text{LP}}^{\eta}E_\text{X}^{\eta}}{(E_\text{C}+E_\text{X}^{\eta})\sqrt{(E_\text{C}-E_\text{X}^{\eta})^2+{(\Omega ^\eta)^2}}}.
\label{hopfield}
\end{align}
The evolution of the Hopfield coefficients with cavity detuning for the M$_{2}^{-}$ upper polariton branch and $\sigma^+$ and $\sigma^-$ polarization are shown in the left and right lower panels of Fig.~\ref{SMfig_hopfield}, respectively. In the case of the upper (lower) polariton branch, $|\mathcal{X}^\eta|^2$ and $|\mathcal{C}^\eta|^2$ quantify the exciton (photon) and photon (exciton) fraction of the respective circularly-polarized polariton state \cite{Kavokin_2011}.

Next, we discuss the evolution of the polariton energy splitting in magnetic field, $\Delta E_{\text{UP/LP}}(B)$, with circularly-polarized polariton state eigenenergies given by Eq.~\eqref{coupled_oscillator}. The contribution of the bare exciton valley Zeeman splitting $\Delta E_\text{X}(B)$ is given by:
\begin{equation}
    \Delta E_\text{X}(B) = E_{\text{X}}^+(B) -E_{\text{X}}^-(B),
\end{equation}
with the magneto-dispersion of the circularly-polarized exciton Zeeman branches determined by the exciton Land\'e factor $g_\text{X}$ as:
\begin{align}
    \label{excenrgy} 
    E_\text{X}^\eta (B) &= E_0 + \eta\frac{g_\text{X}\mu_{B}B}{2},
   \end{align}
where $E_0$ is the exciton energy at zero magnetic field and $\mu_{B}$ is the Bohr magneton. In the absence of correlation-induced magnetism, the magneto-dispersion of the circularly-polarized exciton branches is linear in magnetic field according to the valley Zeeman effect and the exciton $g$-factor \cite{Wang2015}. In the presence of magnetism on electron or hole moir\'e lattices, in contrast, the polaron magneto-dispersions exhibit highly nonlinear evolution with magnetic field \cite{Tang_2020,Xu2022,Campbell_2022,Ciorciaro_2023,Tang_2023}. The contribution of the field-dependent difference in the polarization-contrasting Rabi splittings of polariton branches $\Delta \Omega (B)$, given by
\begin{equation}
    \Delta \Omega (B) = \Omega^+(B) -\Omega^-(B),
    \label{Rabi}
\end{equation}
is polariton-specific and very different in the limits of neutral and charged moir\'e polaritons M$_1$ and M$_2^-$, respectively, as discussed in the following.

With focus on moir\'e M$_1$ exciton-polaritons and M$_2^-$ polaron-polaritons here, we first discuss the evolutions of the energy and oscillator strength of the respective bare M$_1$ moir\'e exciton and M$_2^-$ moir\'e polaron states with magnetic field in the absence of the top DBR mirror. The moir\'e exciton localization sites are shown schematically in Fig.~\ref{SMfig_g-model_motivation}a for the limits of zero doping (top panel) and electron doping with one electron per moir\'e unit cell (bottom panel). We note that due to different localization sites of moir\'e excitons M$_1$ and M$_2$, electron doping at integer filling ($\nu=1$) gives rise to moir\'e polarons M$_1^-$ with co-localized electrons and wavefunctions of the moir\'e exciton M$_1$ (top panel of Fig.~\ref{SMfig1}c), whereas the bulk part of the M$_2$ moir\'e exciton wavefunction (bottom panel of Fig.~\ref{SMfig1}c) resides on a moir\'e site that is distant from the electron localization site, conditioning a distinct charge-doping behavior of the M$_2^-$ polaron \cite{Polovnikov2022,Polovnikov_2023}. Consistently, the response of M$_1^-$ and M$_2^-$ moir\'e polarons to magnetic field is distinct from neutral moir\'e exciton states M$_1$ and M$_2$, and from each other. As shown in the DR spectra of Fig.~\ref{SMfig_g-model_motivation}b, recorded in a magnetic field of $8$~T without the top DBR mirror, the bare M$_1$ and M$_2$ moir\'e exciton states exhibit the characteristic valley Zeeman splitting with equal oscillator strengths of $\sigma^+$ and $\sigma^-$ polarized branches (top panel of Fig.~\ref{SMfig_g-model_motivation}b). For the moir\'e exciton M$_1$, the corresponding $g$-factor is determined by the linear slope of the Zeeman splitting between the peak energies of the two polarized branches as $g_{\text{X}} = -4.2 \pm 0.1$ (top panel of Fig.~\ref{SMfig_g-model_motivation}c), and the change in the oscillator strength between the two Zeeman-split branches $\Delta f$ (normalized by the zero-field value $f_0$) is zero (top panel of Fig.~\ref{SMfig_g-model_motivation}d).      

\begin{figure*}[t!]  
\includegraphics[scale=1.03]{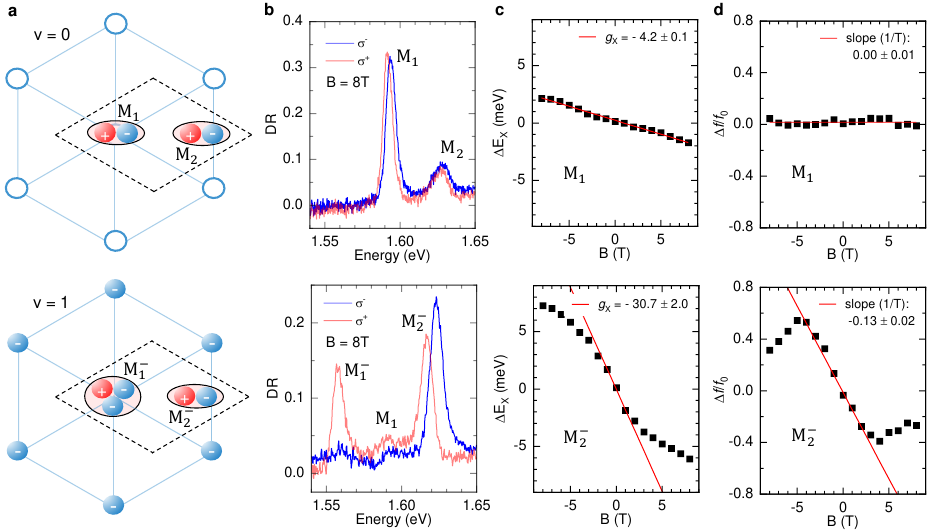}
\vspace{-8pt}
\caption{\textbf{Polarization-contrasting valley Zeeman splittings and oscillator strengths of moir\'e excitons and polarons.} \textbf{a}, Top view schematics of the moir\'e lattice with localization sites of electrons, neutral moir\'e excitons M$_1$ and M$_2$ and moir\'e polarons M$_1^-$ and M$_2^-$ within the moir\'e unit cell for electron filling $\nu = 0$ (upper panel) and $\nu = 1$ (lower panel). \textbf{b}, Polarization-contrasting DR spectra of neutral moir\'e excitons M$_1$ and M$_2$ (upper panel) and electron-dressed moir\'e polarons M$_1^-$ and M$_2^-$ ($\nu =1$, lower panel) at $8$~T in confocal spectroscopy without top DBR mirror. In the neutral regime, M$_1$ and M$_2$ moir\'e peaks exhibit equal oscillator strength for both polarizations. In contrast, at electron filling $\nu =1$, the attractive polaron M$_{1}^{-}$ around $1.56$~eV is nearly completely polarized, whereas the polarization is incomplete for the M$_{2}^{-}$ polaron resonances around $1.62$~eV. The Zeeman-split resonances around $1.59$~eV belong to the repulsive polaron of M$_1$. \textbf{c}, Zeeman splitting of M$_1$ moir\'e excitons (upper panel) and M$_{2}^{-}$ moir\'e polarons (lower panel) as a function of the magnetic field, with respective $g$-factors determined from linear slopes (red solid lines). \textbf{d}, Evolution of the oscillator strength of M$_1$ moir\'e excitons (upper panel) and M$_{2}^{-}$ moir\'e polarons with magnetic field, shown as the field-induced change of the oscillator strength $\Delta f$ normalized by the oscillator strength at zero field $f_0$.}
\label{SMfig_g-model_motivation}
\end{figure*}

The bare M$_1^-$ and M$_2^-$ polarons, on the other hand, exhibit drastically different characteristics in the presence of magnetic field. First, we note that the $\sigma^-$ polarized M$_1^-$ branch is nearly completely quenched at $8$~T, whereas both M$_2^-$ peaks retain finite oscillator strength in both $\sigma^-$ and $\sigma^+$ polarization (bottom panel of Fig.~\ref{SMfig_g-model_motivation}b). This stark difference is consistent with the strong Pauli blocking of the $\sigma^-$ polarized Zeeman branch of the attractive polaron M$_1^-$ forming by co-localization of the M$_1$ moir\'e exciton wavefunction with lattice electrons at $\nu=1$ \cite{Ciorciaro_2023}. The nearly complete polarization of the M$_1^-$ polaron is thus reminiscent of the paramagnetic response of polarons in MoSe$_2$ monolayers due to valley polarization of the electron Fermi-sea \cite{Back_2017}. The M$_2^-$ polaron with marginal electron co-localization, in contrast, shows much reduced quenching. For this reason, the evolution of both Zeeman branches of the M$_2^-$ polaron can be traced up to highest positive and negative magnetic fields of $\pm 8$~T, as shown for the Zeeman splitting and the change in the oscillator strength in the bottom panels of Fig.~\ref{SMfig_g-model_motivation}c and d, respectively: The respective $g$-factor exhibits the nonlinear behavior characteristic of correlation-induced magnetism with $g_{\text{X}}=-30.7 \pm 2.0$ at low magnetic fields, and the difference in the oscillator strength of $\sigma^+$ and $\sigma^-$ polarized Zeeman branches changes by moderate $40\%$ between zero and $3$~T. 

In the framework of M$_1$ and M$_2^-$ polaritons, any change in the oscillator strength with magnetic field will result in polarization-contrasting Rabi splittings $\Delta \Omega (B)$ introduced in Eq.~\eqref{Rabi}. From the analysis in Fig.~\ref{SMfig_g-model_motivation}d, we expect a field-independent Rabi splitting between the two polarized M$_1$ polariton branches, whereas the field-induced change in the oscillator strength of the bare M$_2^-$ polaron will result in a field-dependent Rabi splitting for the $\sigma^+$ and $\sigma^-$ polarized upper and lower M$_2^-$ polaron-polariton branches. At sufficiently small magnetic fields, the polarization of the electron-spin lattice can be approximated by a linear function \cite{Kittel_2004}, and the resulting linearized polarization-contrasting Rabi splitting $\Delta \Omega (B)$ is approximated by
\begin{equation}
    g_{\Omega} = \frac{\Delta \Omega}{\mu_{B} B} = \frac{\Omega^{+} - \Omega^{-}}{\mu_{B} B},
\label{tfactor}
\end{equation}
with the dimensionless parameter $g_{\Omega}$ introduced in analogy to the exciton $g$-factor, and $\Omega^\eta$ given by
\begin{align}
\Omega^\eta (B) &= \Omega_0 +\eta\frac{g_{\Omega}\mu_{B}B}{2},
\label{omega}
\end{align}
with the zero-field Rabi splitting $\Omega_0$. 

\begin{figure*}[t!]  
\includegraphics[scale=1.03]{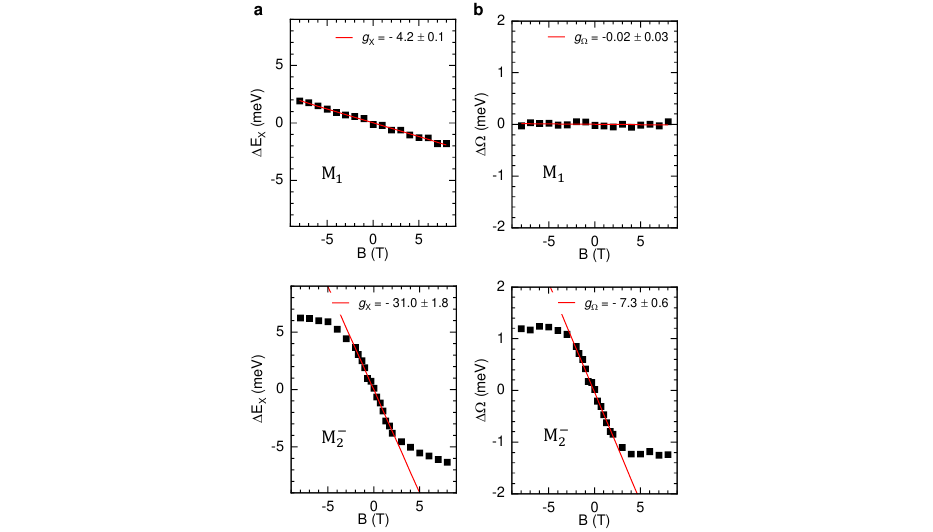}
\vspace{-8pt}
\caption{\textbf{Polarization-contrasting valley Zeeman and Rabi splittings of neutral and charged moir\'e polaritons.} \textbf{a}, Zeeman splitting of M$_1$ moir\'e exciton-polariton (upper panel) and M$_{2}^{-}$ moir\'e polaron-polariton (lower panel) as a function of the magnetic field. \textbf{b}, Change in the Rabi splitting with magnetic field for $\sigma ^+$ and $\sigma ^-$ polarized branches of the M$_1$ exciton-polariton (upper panel) and M$_{2}^{-}$ polaron-polariton. The exciton energies and Rabi splittings in \textbf{a} and \textbf{b} were obtained from fitting Eq.~\eqref{coupled_oscillator} to the respective data.}
\label{SMfig_g-model_polariton}
\end{figure*}

The consequences of the field-independent and field-dependent oscillator strengths of the M$_1$ moir\'e exciton and the M$_2^-$ polaron for the respective polariton Rabi splittings are shown in Fig.~\ref{SMfig_g-model_polariton}b. From the analysis of the polarization-resolved reflection contrast of M$_1$ exciton-polaritons and M$_2^-$ polaron-polaritons at different magnetic fields and cavity-exciton detunings, we obtained the respective exciton Zeeman and Rabi splittings shown in the top and bottom panels of Fig.~\ref{SMfig_g-model_polariton}a and b, respectively. The evolution of the M$_1$ polariton Zeeman splitting is linear in the magnetic field and the extracted exciton $g$-factor $g_{\text{X}}=-4.2\pm0.1$ (top panel of Fig.~\ref{SMfig_g-model_polariton}a) matches the one observed for the bare moir\'e exciton (top panel of Fig.~\ref{SMfig_g-model_motivation}c). Its Rabi splitting is consistently independent of the magnetic field (top panel of Fig.~\ref{SMfig_g-model_polariton}a) and makes no additional contribution to the effective $g$-factor with $g_\Omega=-0.02\pm0.03$. For the M$_2^-$ polariton, in contrast, both the exciton Zeeman and the Rabi splitting are nonlinear in magnetic field (bottom panels of Fig.~\ref{SMfig_g-model_polariton}a and b) as a result of lattice magnetism, with contributions of $g_{\text{X}}=-31.0\pm1.8$ and $g_\Omega=-7.3\pm0.8$ in the small-field limit of $|B| \leq 1$~T.   

With access to both contributions, namely the exciton Zeeman splitting and the polarization-contrasting Rabi splitting, we finally calculate the effective $g$-factor of the upper and lower polariton $g_{\text{UP/LP}}$ by inserting $E_\text{X}^\eta (B)$ and $\Omega^\eta (B)$ from Eqs.~\eqref{excenrgy} and \eqref{omega} into Eq.~\eqref{coupled_oscillator}. In the general case this yields: 
\begin{equation}
    g_{\text{UP/LP}} = \frac{\Delta E_{\text{UP/LP}}}{\mu_{B} B} = \frac{E_{\text{UP/LP}}^{+} -E_{\text{UP/LP}}^{-}}{\mu_{B} B},
\end{equation}
and in the linearized low-field limit of Eqs.~\eqref{tfactor} and \eqref{omega} we derive:
\begin{equation}
    g_{\text{UP/LP}} = 
    \frac{g_\text{X}}{2} \pm \frac{g_\text{X}\delta _{\text{0}} + g_{\Omega} \Omega_{0}}{2\sqrt{\delta _{\text{0}}^2 + \Omega_{0}^{2}}} +\mathcal{O}(B^{2}),
\label{g_model}
\end{equation}
with the zero-field cavity-exciton detuning $\delta _{\text{0}} = E_\text{C}- E_0 $.

\begin{figure*}[t!]  
\includegraphics[scale=1.1]{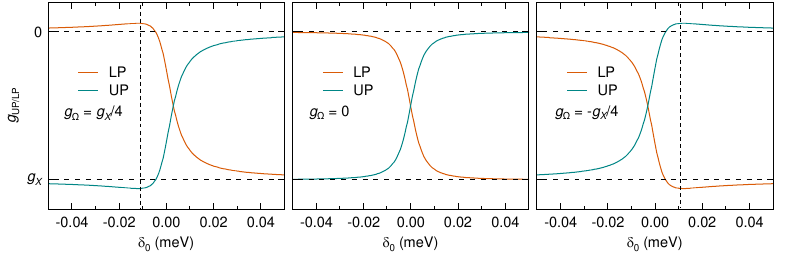}
\vspace{-8pt}
\caption{\textbf{Effective polariton Land\'e factors as a function of the cavity detuning.} Upper and lower polariton $g$-factors $g_{\text{UP/LP}}$, calculated using Eq.~\eqref{g_model} for a negative exciton $g$-factor $g_\text{X}$ and $g_{\Omega}$ = $g_\text{X}/4$ (left panel), 0 (central panel), and $-g_\text{X}/4$ (right panel), respectively. The vertical dashed lines indicate cavity resonance detunings where the upper and lower polariton $g$-factors exhibit extrema for non-vanishing $g_{\Omega}$.}
\label{SMfig_gfactor_model}
\end{figure*}

For large detunings $\delta _{\text{0}}$, the effective polariton $g$-factors approach the limiting values of 0 and $g_{\text{X}}$, when the polariton becomes predominantly photonic and excitonic, respectively. At zero detuning $\delta _{\text{0}} = 0$, the difference between the effective $g$-factors of the upper and lower polariton branch is determined by $g_{\Omega}$ as $(g_{\text{UP}} - g_{\text{LP}})|_{\delta _{\text{0}} = 0}= g_{\Omega}$. Consequently, field-dependent polarization-contrasting Rabi splitting yields finite $g_{\Omega}$, and this conditions an asymmetric evolution of $g_{\text{UP/LP}}$ as function of the cavity detuning $\delta _{\text{0}}$, as observed for the M$_{2}^{-}$ polaron-polariton in Fig.~3d of the main text, in contrast to the symmetric evolution of the M$_{1}$ exciton-polariton $g$-factor in the respective Fig.~3e. 

Finally, it is instructive to discuss the asymmetry in the evolution of the effective polariton $g$-factor $g_{\text{UP/LP}}$ with the cavity detuning $\delta _{\text{0}}$ on the basis of the model results shown in Fig.~\ref{SMfig_gfactor_model}. Obviously, in magnetic fields within the linear-response condition, finite $g_{\Omega}$ renders the dependence of the polariton $g$-factors asymmetric for negative and positive detunings, with asymmetry depending on whether the signs of $g_{\Omega}$ and the exciton $g$-factor $g_\text{X}$ are equal or opposite. In the former case of $\text{sgn}(g_{\Omega})= \text{sgn}(g_\text{X})=-1$, the upper polariton $g$-factor $g_{\text{UP}}$ is enhanced at intermediate detunings as compared to the bare exciton $g$-factor $g_\text{X}$, before approaching $g_\text{X}$ asymptotically at large detunings. Analogously, $g_{\text{LP}}$ changes its sign at intermediate detunings before approaching $0$ at large $\delta _{\text{0}}$. By solving the equation $\partial g_{\text{UP/LP}}/\partial \delta _{\text{0}} = 0$, we obtain the detuning at which the effective polariton $g$-factors have their extrema $\tilde{g}_{\text{UP/LP}}$. For $\text{sgn}(g_{\Omega})=\text{sgn}(g_\text{X})=-1$, and assuming $g_\text{X} \gg g_{\Omega}$, the respective extremal values are approximately given by
\begin{align}
\tilde{g}_{\text{UP}} &\approx g_\text{X} + \frac{g_{\Omega}^2}{g_\text{X}},\\
\tilde{g}_{\text{LP}} &\approx - \frac{g_{\Omega}^2}{g_\text{X}},
\end{align}
in accord with the analysis shown in Fig.~3d of the main text. 

\clearpage
\vspace{-20pt}
%